\documentclass[%
 aip,
 amsmath,amssymb,
reprint,%
]{revtex4-1}

\usepackage{graphicx}% Include figure files
\usepackage{dcolumn}% Align table columns on decimal point
\usepackage{bm}% bold math
\usepackage[utf8]{inputenc}
\usepackage[T1]{fontenc}
\usepackage{mathptmx}

\usepackage{gensymb}
\usepackage{graphicx} 
\usepackage{amsmath} 
\usepackage{placeins}
\usepackage{booktabs}
\usepackage{multirow}
\usepackage{caption}
\usepackage{subcaption}
\usepackage{textcomp}
\usepackage{hyperref}

\begin{document}

%\title[Ferromagnetic-like moment correlations within clusters of iron oxide nanoflowers]{Ferromagnetic-like moment correlations within clusters of iron oxide nanoflowers}
\title{Supraferromagnetic correlations in clusters of magnetic nanoflowers}
% Force line breaks with \\

\author{P. Bender}
\email{philipp.bender@uni.lu.}
\affiliation{Physics and Materials Science Research Unit, University of Luxembourg, L-1511 Luxembourg, Grand Duchy of Luxembourg.}

\author{D. Honecker}%
\altaffiliation[Currently at ]{Physics and Materials Science Research Unit, University of Luxembourg, L-1511 Luxembourg, Grand Duchy of Luxembourg.}
\affiliation{Institut Laue-Langevin, 38042 Grenoble, France.}

\author{L. Fern\'andez Barqu\'in}
\affiliation{Departamento CITIMAC, Faculty of Science, University of Cantabria, 39005 Santander, Spain.}

\date{\today}% It is always \today, today,

\begin{abstract}
Magnetic nanoflowers are densely packed aggregates of superferromagnetically coupled iron oxide nanocrystallites, which excel during magnetic hyperthermia experiments.
Here, we investigate the nature of the moment coupling within a powder of such nanoflowers using spin-resolved small-angle neutron scattering.
Within the powder the nanoparticles are agglomerated to clusters, and we can show that the moments of neighboring nanoflowers tend to align parallel to each other.
Thus, the whole system resembles a hierarchical magnetic nanostructure consisting of three distinct levels, i.e. (i) the ferrimagnetic nanocrystallites as building blocks, (ii) the superferromagnetic nanoflowers, and (iii) the \textit{supra}ferromagnetic clusters of nanoflowers.
We surmise that such a supraferromagnetic coupling explains the enhanced magnetic hyperthermia performance in case of interacting nanoflowers.
\end{abstract}

\maketitle

%\section{Introduction}

The working principle of magnetic hyperthermia (MHT) is to administer a moderate quantity of magnetic nanoparticles within tumors and to heat them up by applying alternating magnetic fields with clinically acceptable  parameters (i.e. comparatively high frequencies $\gtrapprox 100\,\mathrm{kHz}$ but low amplitudes $\lessapprox 20\,\mathrm{mT}$ \cite{pankhurst2009progress,southern2018commentary}) to kill the tumors.
Additionally, a magneto-mechanical actuation of the embedded particles may disrupt the cytoskeleton and lead to cell death. \cite{zhang2014dynamic,master2016remote}
In physiological environment nanoparticles usually agglomerate, which can significantly modify their magnetic properties compared to the dilute non-interacting case \cite{eberbeck2006aggregation,gutierrez2019aggregation} and which in turn may alter their heating behavior. \cite{di2014magnetic,perigo2015fundamentals,sanz2016silico,dejardin2017effect}
Depending on the characteristics of the individual particles and the field parameters, such a clustering can either improve or impair the MHT performance. \cite{mehdaoui2013increase,sadat2014effect,blanco2015high,andreu2015nano,coral2016effect}
In fact it was observed for so-called nanoflowers, which are densely packed aggregates of iron oxide crystallites, that they excel during MHT experiments compared to the single-crystals \cite{lartigue2012cooperative} and other systems such as magnetosomes. \cite{dutz2014magnetic}
This intriguing result motivated numerous studies regarding synthesis and characterization of such flower-shaped particles. \cite{kostopoulou2014assembly,sakellari2016ferrimagnetic,aaa389388fd94f78a5ec927fbdd41d13,gavilan2017formation,wetegrove2019formation,shaw2019synthesis}
It can be shown that an exchange coupling between the cores leads to a superferromagnetic magnetization state \cite{alonso2010crossover} within the individual nanoflowers, \cite{dutz2016magnetic} but with a significant internal spin disorder caused by the high defect density, e.g. due to the grain boundaries. \cite{dobrich2009grain}
It is speculated that such a disordered state enables an increased excitation of the moments  \cite{bender2018relating,bender2018influence}, similar to other defect-rich particles. \cite{lak2018fe2+}
When introduced into tumors, it is safe to assume that the nanoflowers will agglomerate to clusters, and thus interparticle interactions will be relevant. \cite{kuznetsov2019zero}
In \citet{bender2018dipolar} we could show for homogeneous superparamagnetic nanoparticles a predominance for antiferromagnetic-like moment correlations within particle clusters \textit{via} polarized small-angle neutron scattering (SANS).
In this work we use the same approach to determine the nature of the moment coupling within a powder of iron oxide nanoflowers.

\begin{figure}[b]
\centering
\includegraphics[width=1\columnwidth]{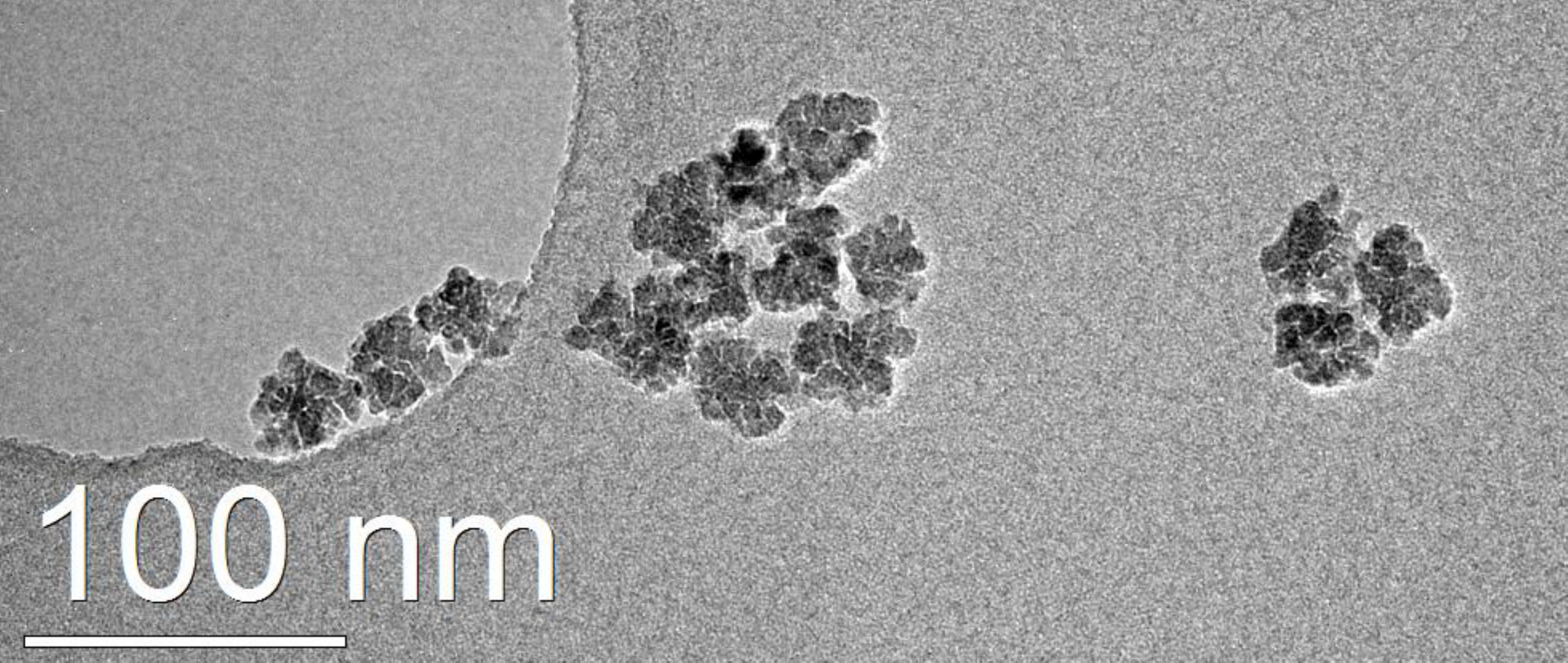}
\caption{TEM image of three separate clusters of nanoflowers. The nuclear SANS results indicate that within the particle powder large clusters with sizes $>160\,\mathrm{nm}$ exist.}
\label{Fig1}
\end{figure}

The synomag-D nanoflowers were supplied by micromod Partikeltechnologie GmbH, Germany, which consist predominately of $\gamma-\mathrm{Fe_2O_3}$, and are coated with dextran.
A detailed study of these particles can be found in \citet{bender2018relating} which showed that they are around 39\,nm in size and consist of crystallites with sizes ranging from $5-15$\,nm.
Transmission electron microscopy (TEM) images were taken with a FEI Titan 80-300 TEM, for which the sample was prepared by putting a small droplet of the dilute dispersion of the particles on a carbon-coated copper grid.
Figure\,\ref{Fig1} shows a typical TEM image of the nanoflowers, in which they are agglomerated to small clusters of 3 and 9 particles, respectively.
As can be seen, the nanoparticles are irregular in shape, and around 30-40\,nm in size.

The polarized SANS experiment of the nanoflower powder \cite{ILLproposal} was performed with the instrument D33 at the Institut Laue-Langevin (ILL), Grenoble (France) \cite{Dewhurst:ks5488} at room-temperature using a mean wavelength of $\lambda=0.6$\,nm ($\Delta\lambda/\lambda=10$\%) and a detector distance of 10.3\,m.
We employed longitudinal neutron-spin analysis (POLARIS) to collect the four spin-resolved intensities $I^{++}(\textbf{q})$, $I^{--}(\textbf{q})$, $I^{+-}(\textbf{q})$ and $I^{-+}(\textbf{q})$, where $+$ denotes the polarization state \textit{spin-up}.
This approach enables the separation of nuclear and magnetic scattering contributions, and was applied in several studies to investigate magnetic nanoparticle ensembles. \cite{krycka2010core,bender2018dipolar,Orue2018}
A homogeneous magnetic field $\textbf{H}$ was applied perpendicular to the neutron beam ($\textbf{H}\perp\textbf{k}$) with a field amplitude of $\mu_0H=2$\,mT, which was necessary to maintain the neutron beam polarization.

\begin{figure*}[ht]
\centering
\includegraphics[width=2\columnwidth]{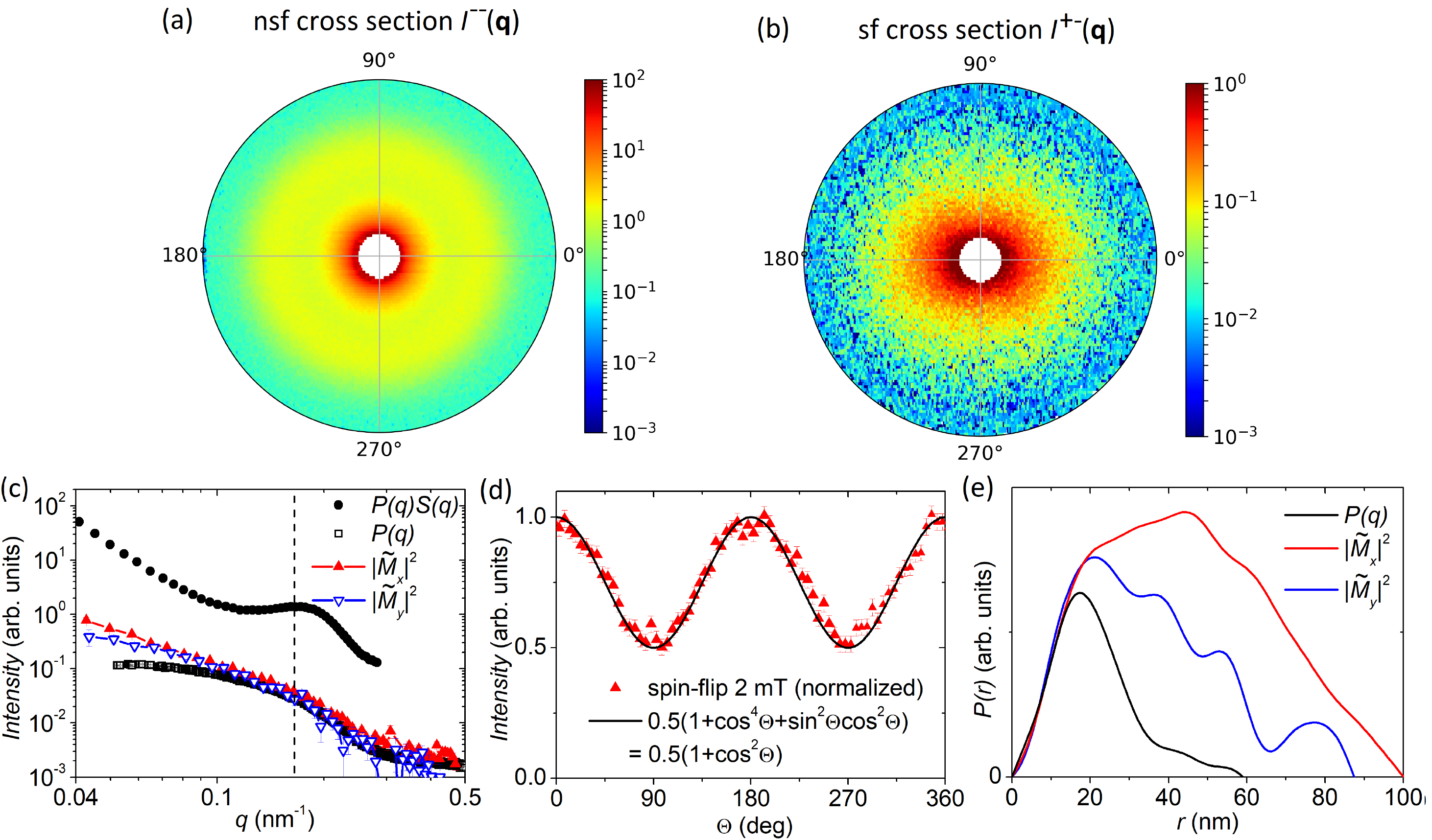}
\caption{Polarized SANS analysis of the nanoflower powder. The magnetic field with $\mu_0H=2\,\mathrm{mT}$ was applied along $\Theta=0^\circ$ and the total accessible $q$-range was around $0.03-0.3\,\mathrm{nm^{-1}}$. (a) 2D scattering pattern of the nsf cross section $I^{--}(\textbf{q})$. 
(b) 2D scattering pattern of the sf cross section $I^{sf}(\textbf{q})=I^{+-}(\textbf{q})$. 
(c) Purely nuclear 1D cross section $I^{\mathrm{nuc}}(q)\propto P(q)S(q)$ extracted from $I^{--}(\textbf{q})$ (sector parallel to \textbf{H}, $\Theta=0^\circ\pm10^\circ$) of the powder, the nuclear cross section $I^{\mathrm{nuc}}(q)\propto P(q)$ determined from $I^{--}(\textbf{q})$ of the dilute colloidal dispersion (from \citet{bender2018relating}), and the magnetic cross sections $|\widetilde{M}_x|^2$ and $|\widetilde{M}_y|^2$ extracted from the sf cross section $I^{\mathrm{sf}}(\textbf{q})$. 
The dashed line at $q=0.17\,\mathrm{nm^{-1}}$ indicates the boarder between the intraparticle length scale (high $q$) and the interparticle length scale (low $q$).
(d) $I^{\mathrm{sf}}(\textbf{q})=I^{+-}(\textbf{q})$ integrated over the whole $q$-range as a function of $\Theta$.
(e) The correlations functions $P(r)$ extracted by indirect Fourier transforms of the 1D nuclear scattering cross section $P(q)$ of the colloid (from \citet{bender2018relating}) and of the magnetic cross sections $|\widetilde{M}_x|^2$ and $|\widetilde{M}_y|^2$.}
\label{Fig2}
\end{figure*}

Figs.\,\ref{Fig2}(a) and (b) display the 2D scattering patterns of the non-spin flip (nsf) cross section $I^{--}(\textbf{q})$ and of the spin-flip (sf) cross section $I^{+-}(\textbf{q})$, respectively.
For the geometry $\mathbf{H}\perp \mathbf{k}$ the nsf cross sections $I^{++}(\textbf{q})$, $I^{--}(\textbf{q})$ can be written as:

\begin{align}\label{Eq1}
I^{\pm\pm}(\mathbf{q})\propto&|\widetilde{N}|^2+b_\mathrm{h}^2|\widetilde{M}_z|^2\mathrm{sin}^4\Theta\nonumber\\
&+b_\mathrm{h}^2|\widetilde{M}_y|^2\mathrm{sin}^2\Theta\mathrm{cos}^2\Theta\nonumber\\
&-b_\mathrm{h}^2(\widetilde{M}_y\widetilde{M}_z^*+\widetilde{M}_z\widetilde{M}_y^*)\mathrm{sin}^3\Theta\mathrm{cos}\Theta\nonumber\\
&\mp b_\mathrm{h}(\widetilde{N}\widetilde{M}_z^*+\widetilde{N}^*\widetilde{M}_z)\mathrm{sin}^2\Theta\nonumber\\
&\pm b_\mathrm{h}(\widetilde{N}\widetilde{M}_y^*+\widetilde{N}^*\widetilde{M}_y)\mathrm{sin}\Theta\mathrm{cos}\Theta,
\end{align}

where $\Theta$ is the angle between the scattering vector $\mathbf{q}=(0,q_y,q_z)$ and the magnetic field $\mathbf{H}$ and $b_\mathrm{h}=2.7\cdot10^{-15}\,\mathrm{m}/\mu_\mathrm{B}$, with $\mu_\mathrm{B}$ being the Bohr magneton. 
Hence, in Figs.\,\ref{Fig2}(a) and (b) the field was applied along $\Theta=0^\circ$.
Moreover, $\tilde{N}(\textbf{q})$ and $\widetilde{\mathbf{M}}=\left[\widetilde{M}_x(\textbf{q}),\widetilde{M}_y(\textbf{q}),\widetilde{M}_z(\textbf{q})\right]$ are the Fourier transforms of the nuclear scattering length density and of the magnetization vector field in the $x$-, $y$- and $z$-directions, respectively, and the index $^*$ denotes the complex conjugate.
One remarkable advantage of POLARIS, is that the purely nuclear scattering can be accessed without further assuming a saturated magnetic system (absence of misaligned moments).
To be precise, the purely nuclear cross section $I^{\mathrm{nuc}}(q)\propto|\tilde{N}|^2$  can be determined, in case of isotropic structures, from the sector parallel to $\mathbf{H}$ of the nsf intensities.

The sf intensities, on the other hand, are of purely magnetic origin.
We assume for our sample that chiral scattering terms can be neglected \cite{muhlbauer2019magnetic}, and thus we can write $I^{\mathrm{sf}}(\textbf{q})=I^{+-}(\textbf{q})=I^{-+}(\textbf{q})$, with (for $\mathbf{H}\perp \mathbf{k}$): \cite{honecker2010longitudinal}
\begin{align}\label{Eq2}
I^{\mathrm{sf}}(\mathbf{q})\propto&|\widetilde{M}_x|^2+|\widetilde{M}_y|^2\mathrm{cos}^4\Theta+|\widetilde{M}_z|^2\mathrm{sin}^2\Theta\mathrm{cos}^2\Theta\nonumber\\
&-(\widetilde{M}_y\widetilde{M}_z^*+\widetilde{M}_z\widetilde{M}_y^*)\mathrm{sin}\Theta\mathrm{cos}^3\Theta.
\end{align} 

The nsf intensity in Fig.\,\ref{Fig2}(a) exhibits basically no anisotropy, indicating the dominance of the isotropic nuclear scattering, and thus verifying a randomly oriented microstructure.
The purely nuclear 1D cross section $I^{\mathrm{nuc}}(q)$ is determined from the sector parallel to magnetic field of $I^{--}(\textbf{q})$ and is plotted in Fig.\,\ref{Fig2}(c).
As can be seen, $I^{\mathrm{nuc}}(q)$ exhibits a peak at around $q=0.17\,\mathrm{nm^{-1}}$.
For particle ensembles the total nuclear cross section is usually written as $I^{\mathrm{nuc}}(q)\propto P(q)S(q)$, where $P(q)$ is the particle form factor and $S(q)$ the structure factor arising from the particle arrangement. \cite{pedersen1997}
For comparison, in Fig.\,\ref{Fig2}(c) we also plot the purely nuclear scattering cross section of the same nanoflowers in dilute colloidal dispersion from \citet{bender2018relating}.
In this case there is no significant structure formation and thus $I^{\mathrm{nuc}}(q)\propto P(q)$.
The observed peak for $I^{\mathrm{nuc}}(q)$ of the powder can be thus attributed to inter-particle correlations, and which implies an average center-to-center distance between the nanoflowers of $2\pi/0.17\,\mathrm{nm^{-1}}=36\,\mathrm{nm}$ (nearest neighbor correlations) \cite{li2016small,alba2016magnetic}.
This estimation is in good agreement with our previous analysis in \citet{bender2018relating}, where we determined an average particle size of around 39\,nm.
For $q\rightarrow 0$ (i.e. the interparticle length scale) the forward scattering intensity increases, which indicates the presence of larger structures within the samples. \cite{li2016small}
Thus we can conclude that no long-range order exists but that the nanoflowers within the powder are agglomerated to large clusters with average sizes outside the minimal $q$-resolution, i.e. average cluster sizes of $>160\,\mathrm{nm}$.

The sf intensity $I^{\mathrm{sf}}(\textbf{q})$ in Fig.\,\ref{Fig2}(b) exhibits a well-pronounced anisotropy, and in Fig.\,\ref{Fig2}(d) we plot $I^{\mathrm{sf}}(q)$ integrated over the whole $q$-range as a function of $\Theta$.
The functional form is well described by the trigonometric terms from Eq.\,\ref{Eq2} without the linear term, which implies equal magnetization along the $x$-, $y$- and $z$-direction and a zero net magnetization. 
This is expected because the sample was in the demagnetized state (i.e., the powder was not exposed to a magnetic field prior to the polarized SANS experiment) and 2\,mT is not sufficient to significantly align the moments (as a reminder, the low magnetic field had to be applied to remain the polarization of the neutron beam).
In Fig.\,\ref{Fig2}(c) we plot $I^{\mathrm{sf}}(\textbf{q})$ determined perpendicular to the field direction, i.e. $I^{\mathrm{sf}}(q, \Theta=90^\circ)\propto|\widetilde{M}_x|^2$, and the difference between $I^{\mathrm{sf}}(q, \Theta=90^\circ)$ and $I^{\mathrm{sf}}(q, \Theta=0^\circ)$, i.e. $|\widetilde{M}_y|^2$.
Both cross sections are in the high $q$-range (i.e. the intraparticle $q$-range) basically identical to each other and to the nuclear particle form factor $P(q)$.
This confirms the superferromagnetic magnetization state within the individual nanoflowers.
In the interparticle $q$-range ($q<0.17\,\,\mathrm{nm^{-1}}$), however, both $|\widetilde{M}_x|^2$ and $|\widetilde{M}_y|^2$ start to deviate from $P(q)$ and increase strongly with decreasing $q$.
Additionally it can be observed in Fig.\,\ref{Fig2}(c) that in the low $q$-range, $|\widetilde{M}_x|^2$ significantly deviates from $|\widetilde{M}_y|^2$.
This can be attributed to the anisotropy of the magnetic structure factor and indicates a disordered microstructure without a short range pseudo-crystalline order. \cite{honecker2019magnetic}
The deviation of both magnetic contributions $|\widetilde{M}_x|^2$ and $|\widetilde{M}_y|^2$ from $P(q)$  is an evidence for interparticle moment correlations between neighboring nanoflowers.
To reveal the nature of these interactions we extracted the underlying magnetic correlation functions $P(r)$ from the scattering intensities by indirect Fourier transforms. \cite{bender2017structural} 
As can be seen in Fig.\,\ref{Fig2}(e), for the two magnetic contributions $|\widetilde{M}_x|^2$ and $|\widetilde{M}_y|^2$ we obtain positive values for $P(r)$ for length scales well above the nanoflower size ($r\geq 36\,\mathrm{nm}$), which indicates positive correlations between the moments of neighboring nanoflowers.
This can be interpreted as evidence for a \textit{supra}ferromagnetic magnetization state within the clusters of these superferromagnetic nanoflowers.

%\section{Conclusions}

We performed a spin-resolved SANS study on a powder of iron oxide nanoflowers, which enables the separation of nuclear and magnetic scattering contributions.
Analysis of the nuclear SANS data shows that the nanoflowers are agglomerated to large clusters.
The magnetic scattering contributions then indicate that the moments between neighboring particles are preferentially aligned parallel to each other.
We interpret this as evidence for a \textit{supra}ferromagnetic magnetization state within the clusters of nanoflowers.
Considering that the nanoflowers itself are aggregates of superferromagnetically coupled crystallites, the whole system can be thus regarded as a hierarchical magnetic nanostructure consisting of three distinct levels, i.e. (i) the ferrimagnetic nanocrystallites as building blocks, (ii) the superferromagnetic nanoflowers, and (iii) the supraferromagnetic clusters of nanoflowers.
It can be assumed that such supraferromagnetic correlations increase the low-field susceptibility of the ensemble and thus its MHT performance compared to the dilute, non-interacting ensemble. 
Indeed, we surmise that our observation explains the intriguing result in \citet{sakellari2016ferrimagnetic} where for colloidal dispersions of 50-nm nanoflowers an increased heating with increasing particle concentration was detected, which is in contrast to other nanoparticle ensembles for which usually increasing interactions result in a decrease of the MHT performance. \cite{urtizberea2010specific,serantes2010influence}
Considering that in physiological environments usually a clustering of immersed nanoparticles occurs it is a promising result for such nanoflowers that their exceptional heating behavior can be even further enhanced by cluster formation.
For further studies we propose a systematic investigation of the relations between cluster size and MHT performance for embedded nanoflowers, ideally accompanied by polarized SANS studies to probe the interparticle moment correlations.

%\section*{Acknowledgements}

We want to thank the Institut Laue-Langevin for provision of beamtime at the instrument D33 and David Gonz\'alez-Alonso for his help during the experiment.
This project has received funding from the European Commission Framework Programme 7 under Grant Agreement No. 604448 (NanoMag), the National Research Fund of Luxembourg (CORE SANS4NCC grant) and the Spanish Government (MAT2017-83631-C3-R).

\nocite{*}
\bibliography{PBenderBib} %your .bib file

\begin{thebibliography}{44}%
\makeatletter
\providecommand \@ifxundefined [1]{%
 \@ifx{#1\undefined}
}%
\providecommand \@ifnum [1]{%
 \ifnum #1\expandafter \@firstoftwo
 \else \expandafter \@secondoftwo
 \fi
}%
\providecommand \@ifx [1]{%
 \ifx #1\expandafter \@firstoftwo
 \else \expandafter \@secondoftwo
 \fi
}%
\providecommand \natexlab [1]{#1}%
\providecommand \enquote  [1]{``#1''}%
\providecommand \bibnamefont  [1]{#1}%
\providecommand \bibfnamefont [1]{#1}%
\providecommand \citenamefont [1]{#1}%
\providecommand \href@noop [0]{\@secondoftwo}%
\providecommand \href [0]{\begingroup \@sanitize@url \@href}%
\providecommand \@href[1]{\@@startlink{#1}\@@href}%
\providecommand \@@href[1]{\endgroup#1\@@endlink}%
\providecommand \@sanitize@url [0]{\catcode `\\12\catcode `\$12\catcode
  `\&12\catcode `\#12\catcode `\^12\catcode `\_12\catcode `\%12\relax}%
\providecommand \@@startlink[1]{}%
\providecommand \@@endlink[0]{}%
\providecommand \url  [0]{\begingroup\@sanitize@url \@url }%
\providecommand \@url [1]{\endgroup\@href {#1}{\urlprefix }}%
\providecommand \urlprefix  [0]{URL }%
\providecommand \Eprint [0]{\href }%
\providecommand \doibase [0]{http://dx.doi.org/}%
\providecommand \selectlanguage [0]{\@gobble}%
\providecommand \bibinfo  [0]{\@secondoftwo}%
\providecommand \bibfield  [0]{\@secondoftwo}%
\providecommand \translation [1]{[#1]}%
\providecommand \BibitemOpen [0]{}%
\providecommand \bibitemStop [0]{}%
\providecommand \bibitemNoStop [0]{.\EOS\space}%
\providecommand \EOS [0]{\spacefactor3000\relax}%
\providecommand \BibitemShut  [1]{\csname bibitem#1\endcsname}%
\let\auto@bib@innerbib\@empty
%</preamble>
\bibitem [{\citenamefont {Pankhurst}\ \emph {et~al.}(2009)\citenamefont
  {Pankhurst}, \citenamefont {Thanh}, \citenamefont {Jones},\ and\
  \citenamefont {Dobson}}]{pankhurst2009progress}%
  \BibitemOpen
  \bibfield  {author} {\bibinfo {author} {\bibfnamefont {Q.}~\bibnamefont
  {Pankhurst}}, \bibinfo {author} {\bibfnamefont {N.}~\bibnamefont {Thanh}},
  \bibinfo {author} {\bibfnamefont {S.}~\bibnamefont {Jones}}, \ and\ \bibinfo
  {author} {\bibfnamefont {J.}~\bibnamefont {Dobson}},\ }\bibfield  {title}
  {\enquote {\bibinfo {title} {Progress in applications of magnetic
  nanoparticles in biomedicine},}\ }\href {\doibase
  10.1088/0022-3727/42/22/224001} {\bibfield  {journal} {\bibinfo  {journal}
  {J. Phys. D: Appl. Phys.}\ }\textbf {\bibinfo {volume} {42}},\ \bibinfo
  {pages} {224001} (\bibinfo {year} {2009})}\BibitemShut {NoStop}%
\bibitem [{\citenamefont {Southern}\ and\ \citenamefont
  {Pankhurst}(2018)}]{southern2018commentary}%
  \BibitemOpen
  \bibfield  {author} {\bibinfo {author} {\bibfnamefont {P.}~\bibnamefont
  {Southern}}\ and\ \bibinfo {author} {\bibfnamefont {Q.~A.}\ \bibnamefont
  {Pankhurst}},\ }\bibfield  {title} {\enquote {\bibinfo {title} {Commentary on
  the clinical and preclinical dosage limits of interstitially administered
  magnetic fluids for therapeutic hyperthermia based on current practice and
  efficacy models},}\ }\href {\doibase 10.1080/02656736.2017.1365953}
  {\bibfield  {journal} {\bibinfo  {journal} {Int. J. Hyperth.}\ }\textbf
  {\bibinfo {volume} {34}},\ \bibinfo {pages} {671--686} (\bibinfo {year}
  {2018})}\BibitemShut {NoStop}%
\bibitem [{\citenamefont {Zhang}\ \emph {et~al.}(2014)\citenamefont {Zhang},
  \citenamefont {Kircher}, \citenamefont {Koch}, \citenamefont {Eliasson},
  \citenamefont {Goldberg},\ and\ \citenamefont
  {Renstr\"om}}]{zhang2014dynamic}%
  \BibitemOpen
  \bibfield  {author} {\bibinfo {author} {\bibfnamefont {E.}~\bibnamefont
  {Zhang}}, \bibinfo {author} {\bibfnamefont {M.~F.}\ \bibnamefont {Kircher}},
  \bibinfo {author} {\bibfnamefont {M.}~\bibnamefont {Koch}}, \bibinfo {author}
  {\bibfnamefont {L.}~\bibnamefont {Eliasson}}, \bibinfo {author}
  {\bibfnamefont {S.~N.}\ \bibnamefont {Goldberg}}, \ and\ \bibinfo {author}
  {\bibfnamefont {E.}~\bibnamefont {Renstr\"om}},\ }\bibfield  {title}
  {\enquote {\bibinfo {title} {Dynamic magnetic fields remote-control apoptosis
  via nanoparticle rotation},}\ }\href {\doibase 10.1021/nn406302j} {\bibfield
  {journal} {\bibinfo  {journal} {ACS Nano}\ }\textbf {\bibinfo {volume} {8}},\
  \bibinfo {pages} {3192--3201} (\bibinfo {year} {2014})}\BibitemShut {NoStop}%
\bibitem [{\citenamefont {Master}\ \emph {et~al.}(2016)\citenamefont {Master},
  \citenamefont {Williams}, \citenamefont {Pothayee}, \citenamefont {Pothayee},
  \citenamefont {Zhang}, \citenamefont {Vishwasrao}, \citenamefont {Golovin},
  \citenamefont {Riffle}, \citenamefont {Sokolsky},\ and\ \citenamefont
  {Kabanov}}]{master2016remote}%
  \BibitemOpen
  \bibfield  {author} {\bibinfo {author} {\bibfnamefont {A.~M.}\ \bibnamefont
  {Master}}, \bibinfo {author} {\bibfnamefont {P.~N.}\ \bibnamefont
  {Williams}}, \bibinfo {author} {\bibfnamefont {N.}~\bibnamefont {Pothayee}},
  \bibinfo {author} {\bibfnamefont {N.}~\bibnamefont {Pothayee}}, \bibinfo
  {author} {\bibfnamefont {R.}~\bibnamefont {Zhang}}, \bibinfo {author}
  {\bibfnamefont {H.~M.}\ \bibnamefont {Vishwasrao}}, \bibinfo {author}
  {\bibfnamefont {Y.~I.}\ \bibnamefont {Golovin}}, \bibinfo {author}
  {\bibfnamefont {J.~S.}\ \bibnamefont {Riffle}}, \bibinfo {author}
  {\bibfnamefont {M.}~\bibnamefont {Sokolsky}}, \ and\ \bibinfo {author}
  {\bibfnamefont {A.~V.}\ \bibnamefont {Kabanov}},\ }\bibfield  {title}
  {\enquote {\bibinfo {title} {Remote actuation of magnetic nanoparticles for
  cancer cell selective treatment through cytoskeletal disruption},}\ }\href
  {\doibase 10.1038/srep33560} {\bibfield  {journal} {\bibinfo  {journal} {Sci.
  Rep.}\ }\textbf {\bibinfo {volume} {6}},\ \bibinfo {pages} {33560} (\bibinfo
  {year} {2016})}\BibitemShut {NoStop}%
\bibitem [{\citenamefont {Eberbeck}\ \emph {et~al.}(2006)\citenamefont
  {Eberbeck}, \citenamefont {Wiekhorst}, \citenamefont {Steinhoff},\ and\
  \citenamefont {Trahms}}]{eberbeck2006aggregation}%
  \BibitemOpen
  \bibfield  {author} {\bibinfo {author} {\bibfnamefont {D.}~\bibnamefont
  {Eberbeck}}, \bibinfo {author} {\bibfnamefont {F.}~\bibnamefont {Wiekhorst}},
  \bibinfo {author} {\bibfnamefont {U.}~\bibnamefont {Steinhoff}}, \ and\
  \bibinfo {author} {\bibfnamefont {L.}~\bibnamefont {Trahms}},\ }\bibfield
  {title} {\enquote {\bibinfo {title} {Aggregation behaviour of magnetic
  nanoparticle suspensions investigated by magnetorelaxometry},}\ }\href
  {\doibase 10.1088/0953-8984/18/38/S20} {\bibfield  {journal} {\bibinfo
  {journal} {J. Phys.: Condens. Matter}\ }\textbf {\bibinfo {volume} {18}},\
  \bibinfo {pages} {S2829} (\bibinfo {year} {2006})}\BibitemShut {NoStop}%
\bibitem [{\citenamefont {Gutierrez}\ \emph {et~al.}(2019)\citenamefont
  {Gutierrez}, \citenamefont {Moros}, \citenamefont {Mazario}, \citenamefont
  {de~Bernardo}, \citenamefont {de~la Fuente}, \citenamefont {del
  Puerto~Morales},\ and\ \citenamefont {Salas}}]{gutierrez2019aggregation}%
  \BibitemOpen
  \bibfield  {author} {\bibinfo {author} {\bibfnamefont {L.}~\bibnamefont
  {Gutierrez}}, \bibinfo {author} {\bibfnamefont {M.}~\bibnamefont {Moros}},
  \bibinfo {author} {\bibfnamefont {E.}~\bibnamefont {Mazario}}, \bibinfo
  {author} {\bibfnamefont {S.}~\bibnamefont {de~Bernardo}}, \bibinfo {author}
  {\bibfnamefont {J.~M.}\ \bibnamefont {de~la Fuente}}, \bibinfo {author}
  {\bibfnamefont {M.}~\bibnamefont {del Puerto~Morales}}, \ and\ \bibinfo
  {author} {\bibfnamefont {G.}~\bibnamefont {Salas}},\ }\bibfield  {title}
  {\enquote {\bibinfo {title} {Aggregation effects on the magnetic properties
  of iron oxide colloids},}\ }\href {\doibase 10.1088/1361-6528/aafbff}
  {\bibfield  {journal} {\bibinfo  {journal} {Nanotechnology}\ }\textbf
  {\bibinfo {volume} {30}},\ \bibinfo {pages} {112001} (\bibinfo {year}
  {2019})}\BibitemShut {NoStop}%
\bibitem [{\citenamefont {Di~Corato}\ \emph {et~al.}(2014)\citenamefont
  {Di~Corato}, \citenamefont {Espinosa}, \citenamefont {Lartigue},
  \citenamefont {Tharaud}, \citenamefont {Chat}, \citenamefont {Pellegrino},
  \citenamefont {M{\'e}nager}, \citenamefont {Gazeau},\ and\ \citenamefont
  {Wilhelm}}]{di2014magnetic}%
  \BibitemOpen
  \bibfield  {author} {\bibinfo {author} {\bibfnamefont {R.}~\bibnamefont
  {Di~Corato}}, \bibinfo {author} {\bibfnamefont {A.}~\bibnamefont {Espinosa}},
  \bibinfo {author} {\bibfnamefont {L.}~\bibnamefont {Lartigue}}, \bibinfo
  {author} {\bibfnamefont {M.}~\bibnamefont {Tharaud}}, \bibinfo {author}
  {\bibfnamefont {S.}~\bibnamefont {Chat}}, \bibinfo {author} {\bibfnamefont
  {T.}~\bibnamefont {Pellegrino}}, \bibinfo {author} {\bibfnamefont
  {C.}~\bibnamefont {M{\'e}nager}}, \bibinfo {author} {\bibfnamefont
  {F.}~\bibnamefont {Gazeau}}, \ and\ \bibinfo {author} {\bibfnamefont
  {C.}~\bibnamefont {Wilhelm}},\ }\bibfield  {title} {\enquote {\bibinfo
  {title} {Magnetic hyperthermia efficiency in the cellular environment for
  different nanoparticle designs},}\ }\href {\doibase
  10.1016/j.biomaterials.2014.04.036} {\bibfield  {journal} {\bibinfo
  {journal} {Biomaterials}\ }\textbf {\bibinfo {volume} {35}},\ \bibinfo
  {pages} {6400--6411} (\bibinfo {year} {2014})}\BibitemShut {NoStop}%
\bibitem [{\citenamefont {P{\'e}rigo}\ \emph {et~al.}(2015)\citenamefont
  {P{\'e}rigo}, \citenamefont {Hemery}, \citenamefont {Sandre}, \citenamefont
  {Ortega}, \citenamefont {Garaio}, \citenamefont {Plazaola},\ and\
  \citenamefont {Teran}}]{perigo2015fundamentals}%
  \BibitemOpen
  \bibfield  {author} {\bibinfo {author} {\bibfnamefont {E.~A.}\ \bibnamefont
  {P{\'e}rigo}}, \bibinfo {author} {\bibfnamefont {G.}~\bibnamefont {Hemery}},
  \bibinfo {author} {\bibfnamefont {O.}~\bibnamefont {Sandre}}, \bibinfo
  {author} {\bibfnamefont {D.}~\bibnamefont {Ortega}}, \bibinfo {author}
  {\bibfnamefont {E.}~\bibnamefont {Garaio}}, \bibinfo {author} {\bibfnamefont
  {F.}~\bibnamefont {Plazaola}}, \ and\ \bibinfo {author} {\bibfnamefont
  {F.~J.}\ \bibnamefont {Teran}},\ }\bibfield  {title} {\enquote {\bibinfo
  {title} {Fundamentals and advances in magnetic hyperthermia},}\ }\href
  {\doibase 10.1063/1.4935688} {\bibfield  {journal} {\bibinfo  {journal}
  {Appl. Phys. Rev.}\ }\textbf {\bibinfo {volume} {2}},\ \bibinfo {pages}
  {041302} (\bibinfo {year} {2015})}\BibitemShut {NoStop}%
\bibitem [{\citenamefont {Sanz}\ \emph {et~al.}(2016)\citenamefont {Sanz},
  \citenamefont {Calatayud}, \citenamefont {De~Biasi}, \citenamefont {Lima~Jr},
  \citenamefont {Mansilla}, \citenamefont {Zysler}, \citenamefont {Ibarra},\
  and\ \citenamefont {Goya}}]{sanz2016silico}%
  \BibitemOpen
  \bibfield  {author} {\bibinfo {author} {\bibfnamefont {B.}~\bibnamefont
  {Sanz}}, \bibinfo {author} {\bibfnamefont {M.~P.}\ \bibnamefont {Calatayud}},
  \bibinfo {author} {\bibfnamefont {E.}~\bibnamefont {De~Biasi}}, \bibinfo
  {author} {\bibfnamefont {E.}~\bibnamefont {Lima~Jr}}, \bibinfo {author}
  {\bibfnamefont {M.~V.}\ \bibnamefont {Mansilla}}, \bibinfo {author}
  {\bibfnamefont {R.~D.}\ \bibnamefont {Zysler}}, \bibinfo {author}
  {\bibfnamefont {M.~R.}\ \bibnamefont {Ibarra}}, \ and\ \bibinfo {author}
  {\bibfnamefont {G.~F.}\ \bibnamefont {Goya}},\ }\bibfield  {title} {\enquote
  {\bibinfo {title} {In silico before in vivo: how to predict the heating
  efficiency of magnetic nanoparticles within the intracellular space},}\
  }\href {\doibase 10.1038/srep38733} {\bibfield  {journal} {\bibinfo
  {journal} {Sci. Rep.}\ }\textbf {\bibinfo {volume} {6}},\ \bibinfo {pages}
  {38733} (\bibinfo {year} {2016})}\BibitemShut {NoStop}%
\bibitem [{\citenamefont {D{\'e}jardin}\ \emph {et~al.}(2017)\citenamefont
  {D{\'e}jardin}, \citenamefont {Vernay}, \citenamefont {Respaud},\ and\
  \citenamefont {Kachkachi}}]{dejardin2017effect}%
  \BibitemOpen
  \bibfield  {author} {\bibinfo {author} {\bibfnamefont {J.-L.}\ \bibnamefont
  {D{\'e}jardin}}, \bibinfo {author} {\bibfnamefont {F.}~\bibnamefont
  {Vernay}}, \bibinfo {author} {\bibfnamefont {M.}~\bibnamefont {Respaud}}, \
  and\ \bibinfo {author} {\bibfnamefont {H.}~\bibnamefont {Kachkachi}},\
  }\bibfield  {title} {\enquote {\bibinfo {title} {Effect of dipolar
  interactions and dc magnetic field on the specific absorption rate of an
  array of magnetic nanoparticles},}\ }\href {\doibase 10.1063/1.4984013}
  {\bibfield  {journal} {\bibinfo  {journal} {J. Appl. Phys.}\ }\textbf
  {\bibinfo {volume} {121}},\ \bibinfo {pages} {203903} (\bibinfo {year}
  {2017})}\BibitemShut {NoStop}%
\bibitem [{\citenamefont {Mehdaoui}\ \emph {et~al.}(2013)\citenamefont
  {Mehdaoui}, \citenamefont {Tan}, \citenamefont {Meffre}, \citenamefont
  {Carrey}, \citenamefont {Lachaize}, \citenamefont {Chaudret},\ and\
  \citenamefont {Respaud}}]{mehdaoui2013increase}%
  \BibitemOpen
  \bibfield  {author} {\bibinfo {author} {\bibfnamefont {B.}~\bibnamefont
  {Mehdaoui}}, \bibinfo {author} {\bibfnamefont {R.}~\bibnamefont {Tan}},
  \bibinfo {author} {\bibfnamefont {A.}~\bibnamefont {Meffre}}, \bibinfo
  {author} {\bibfnamefont {J.}~\bibnamefont {Carrey}}, \bibinfo {author}
  {\bibfnamefont {S.}~\bibnamefont {Lachaize}}, \bibinfo {author}
  {\bibfnamefont {B.}~\bibnamefont {Chaudret}}, \ and\ \bibinfo {author}
  {\bibfnamefont {M.}~\bibnamefont {Respaud}},\ }\bibfield  {title} {\enquote
  {\bibinfo {title} {Increase of magnetic hyperthermia efficiency due to
  dipolar interactions in low-anisotropy magnetic nanoparticles: Theoretical
  and experimental results},}\ }\href {\doibase 10.1103/PhysRevB.87.174419}
  {\bibfield  {journal} {\bibinfo  {journal} {Phys. Rev. B}\ }\textbf {\bibinfo
  {volume} {87}},\ \bibinfo {pages} {174419} (\bibinfo {year}
  {2013})}\BibitemShut {NoStop}%
\bibitem [{\citenamefont {Sadat}\ \emph {et~al.}(2014)\citenamefont {Sadat},
  \citenamefont {Patel}, \citenamefont {Sookoor}, \citenamefont {Bud'ko},
  \citenamefont {Ewing}, \citenamefont {Zhang}, \citenamefont {Xu},
  \citenamefont {Wang}, \citenamefont {Pauletti}, \citenamefont {Mast},\ and\
  \citenamefont {Shi}}]{sadat2014effect}%
  \BibitemOpen
  \bibfield  {author} {\bibinfo {author} {\bibfnamefont {M.}~\bibnamefont
  {Sadat}}, \bibinfo {author} {\bibfnamefont {R.}~\bibnamefont {Patel}},
  \bibinfo {author} {\bibfnamefont {J.}~\bibnamefont {Sookoor}}, \bibinfo
  {author} {\bibfnamefont {S.~L.}\ \bibnamefont {Bud'ko}}, \bibinfo {author}
  {\bibfnamefont {R.~C.}\ \bibnamefont {Ewing}}, \bibinfo {author}
  {\bibfnamefont {J.}~\bibnamefont {Zhang}}, \bibinfo {author} {\bibfnamefont
  {H.}~\bibnamefont {Xu}}, \bibinfo {author} {\bibfnamefont {Y.}~\bibnamefont
  {Wang}}, \bibinfo {author} {\bibfnamefont {G.~M.}\ \bibnamefont {Pauletti}},
  \bibinfo {author} {\bibfnamefont {D.~B.}\ \bibnamefont {Mast}}, \ and\
  \bibinfo {author} {\bibfnamefont {D.}~\bibnamefont {Shi}},\ }\bibfield
  {title} {\enquote {\bibinfo {title} {Effect of spatial confinement on
  magnetic hyperthermia via dipolar interactions in $\mathrm{Fe_3O_4}$
  nanoparticles for biomedical applications},}\ }\href {\doibase
  10.1016/j.msec.2014.04.064} {\bibfield  {journal} {\bibinfo  {journal}
  {Mater. Sci. Eng., C}\ }\textbf {\bibinfo {volume} {42}},\ \bibinfo {pages}
  {52--63} (\bibinfo {year} {2014})}\BibitemShut {NoStop}%
\bibitem [{\citenamefont {Blanco-Andujar}\ \emph {et~al.}(2015)\citenamefont
  {Blanco-Andujar}, \citenamefont {Ortega}, \citenamefont {Southern},
  \citenamefont {Pankhurst},\ and\ \citenamefont {Thanh}}]{blanco2015high}%
  \BibitemOpen
  \bibfield  {author} {\bibinfo {author} {\bibfnamefont {C.}~\bibnamefont
  {Blanco-Andujar}}, \bibinfo {author} {\bibfnamefont {D.}~\bibnamefont
  {Ortega}}, \bibinfo {author} {\bibfnamefont {P.}~\bibnamefont {Southern}},
  \bibinfo {author} {\bibfnamefont {Q.}~\bibnamefont {Pankhurst}}, \ and\
  \bibinfo {author} {\bibfnamefont {N.}~\bibnamefont {Thanh}},\ }\bibfield
  {title} {\enquote {\bibinfo {title} {High performance multi-core iron oxide
  nanoparticles for magnetic hyperthermia: microwave synthesis, and the role of
  core-to-core interactions},}\ }\href {\doibase 10.1039/C4NR06239F} {\bibfield
   {journal} {\bibinfo  {journal} {Nanoscale}\ }\textbf {\bibinfo {volume}
  {7}},\ \bibinfo {pages} {1768--1775} (\bibinfo {year} {2015})}\BibitemShut
  {NoStop}%
\bibitem [{\citenamefont {Andreu}\ \emph {et~al.}(2015)\citenamefont {Andreu},
  \citenamefont {Natividad}, \citenamefont {Solozabal},\ and\ \citenamefont
  {Roubeau}}]{andreu2015nano}%
  \BibitemOpen
  \bibfield  {author} {\bibinfo {author} {\bibfnamefont {I.}~\bibnamefont
  {Andreu}}, \bibinfo {author} {\bibfnamefont {E.}~\bibnamefont {Natividad}},
  \bibinfo {author} {\bibfnamefont {L.}~\bibnamefont {Solozabal}}, \ and\
  \bibinfo {author} {\bibfnamefont {O.}~\bibnamefont {Roubeau}},\ }\bibfield
  {title} {\enquote {\bibinfo {title} {Nano-objects for addressing the control
  of nanoparticle arrangement and performance in magnetic hyperthermia},}\
  }\href {\doibase 10.1021/nn505781f} {\bibfield  {journal} {\bibinfo
  {journal} {ACS Nano}\ }\textbf {\bibinfo {volume} {9}},\ \bibinfo {pages}
  {1408--1419} (\bibinfo {year} {2015})}\BibitemShut {NoStop}%
\bibitem [{\citenamefont {Coral}\ \emph {et~al.}(2016)\citenamefont {Coral},
  \citenamefont {Mendoza~Z\'elis}, \citenamefont {Marciello}, \citenamefont
  {Morales}, \citenamefont {Craievich}, \citenamefont {S\'anchez},\ and\
  \citenamefont {Fern\'andez~van Raap}}]{coral2016effect}%
  \BibitemOpen
  \bibfield  {author} {\bibinfo {author} {\bibfnamefont {D.~F.}\ \bibnamefont
  {Coral}}, \bibinfo {author} {\bibfnamefont {P.}~\bibnamefont
  {Mendoza~Z\'elis}}, \bibinfo {author} {\bibfnamefont {M.}~\bibnamefont
  {Marciello}}, \bibinfo {author} {\bibfnamefont {M.~d.~P.}\ \bibnamefont
  {Morales}}, \bibinfo {author} {\bibfnamefont {A.}~\bibnamefont {Craievich}},
  \bibinfo {author} {\bibfnamefont {F.~H.}\ \bibnamefont {S\'anchez}}, \ and\
  \bibinfo {author} {\bibfnamefont {M.~B.}\ \bibnamefont {Fern\'andez~van
  Raap}},\ }\bibfield  {title} {\enquote {\bibinfo {title} {Effect of
  nanoclustering and dipolar interactions in heat generation for magnetic
  hyperthermia},}\ }\href {\doibase 10.1021/acs.langmuir.5b03559} {\bibfield
  {journal} {\bibinfo  {journal} {Langmuir}\ }\textbf {\bibinfo {volume}
  {32}},\ \bibinfo {pages} {1201--1213} (\bibinfo {year} {2016})}\BibitemShut
  {NoStop}%
\bibitem [{\citenamefont {Lartigue}\ \emph {et~al.}(2012)\citenamefont
  {Lartigue}, \citenamefont {Hugounenq}, \citenamefont {Alloyeau},
  \citenamefont {Clarke}, \citenamefont {L\'{e}vy}, \citenamefont {Bacri},
  \citenamefont {Bazzi}, \citenamefont {Brougham}, \citenamefont {Wilhelm},\
  and\ \citenamefont {Gazeau}}]{lartigue2012cooperative}%
  \BibitemOpen
  \bibfield  {author} {\bibinfo {author} {\bibfnamefont {L.}~\bibnamefont
  {Lartigue}}, \bibinfo {author} {\bibfnamefont {P.}~\bibnamefont {Hugounenq}},
  \bibinfo {author} {\bibfnamefont {D.}~\bibnamefont {Alloyeau}}, \bibinfo
  {author} {\bibfnamefont {S.~P.}\ \bibnamefont {Clarke}}, \bibinfo {author}
  {\bibfnamefont {M.}~\bibnamefont {L\'{e}vy}}, \bibinfo {author}
  {\bibfnamefont {J.-C.}\ \bibnamefont {Bacri}}, \bibinfo {author}
  {\bibfnamefont {R.}~\bibnamefont {Bazzi}}, \bibinfo {author} {\bibfnamefont
  {D.~F.}\ \bibnamefont {Brougham}}, \bibinfo {author} {\bibfnamefont
  {C.}~\bibnamefont {Wilhelm}}, \ and\ \bibinfo {author} {\bibfnamefont
  {F.}~\bibnamefont {Gazeau}},\ }\bibfield  {title} {\enquote {\bibinfo {title}
  {Cooperative organization in iron oxide multi-core nanoparticles potentiates
  their efficiency as heating mediators and mri contrast agents},}\ }\href
  {\doibase 10.1021/nn304477s} {\bibfield  {journal} {\bibinfo  {journal} {ACS
  Nano}\ }\textbf {\bibinfo {volume} {6}},\ \bibinfo {pages} {10935--10949}
  (\bibinfo {year} {2012})}\BibitemShut {NoStop}%
\bibitem [{\citenamefont {Dutz}\ and\ \citenamefont
  {Hergt}(2014)}]{dutz2014magnetic}%
  \BibitemOpen
  \bibfield  {author} {\bibinfo {author} {\bibfnamefont {S.}~\bibnamefont
  {Dutz}}\ and\ \bibinfo {author} {\bibfnamefont {R.}~\bibnamefont {Hergt}},\
  }\bibfield  {title} {\enquote {\bibinfo {title} {Magnetic particle
  hyperthermia—a promising tumour therapy?}}\ }\href {\doibase
  10.1088/0957-4484/25/45/452001} {\bibfield  {journal} {\bibinfo  {journal}
  {Nanotechnology}\ }\textbf {\bibinfo {volume} {25}},\ \bibinfo {pages}
  {452001} (\bibinfo {year} {2014})}\BibitemShut {NoStop}%
\bibitem [{\citenamefont {Kostopoulou}\ \emph {et~al.}(2014)\citenamefont
  {Kostopoulou}, \citenamefont {Brintakis}, \citenamefont {Vasilakaki},
  \citenamefont {Trohidou}, \citenamefont {Douvalis}, \citenamefont
  {Lascialfari}, \citenamefont {Manna},\ and\ \citenamefont
  {Lappas}}]{kostopoulou2014assembly}%
  \BibitemOpen
  \bibfield  {author} {\bibinfo {author} {\bibfnamefont {A.}~\bibnamefont
  {Kostopoulou}}, \bibinfo {author} {\bibfnamefont {K.}~\bibnamefont
  {Brintakis}}, \bibinfo {author} {\bibfnamefont {M.}~\bibnamefont
  {Vasilakaki}}, \bibinfo {author} {\bibfnamefont {K.}~\bibnamefont
  {Trohidou}}, \bibinfo {author} {\bibfnamefont {A.}~\bibnamefont {Douvalis}},
  \bibinfo {author} {\bibfnamefont {A.}~\bibnamefont {Lascialfari}}, \bibinfo
  {author} {\bibfnamefont {L.}~\bibnamefont {Manna}}, \ and\ \bibinfo {author}
  {\bibfnamefont {A.}~\bibnamefont {Lappas}},\ }\bibfield  {title} {\enquote
  {\bibinfo {title} {Assembly-mediated interplay of dipolar interactions and
  surface spin disorder in colloidal maghemite nanoclusters},}\ }\href
  {\doibase 10.1039/c3nr06103e} {\bibfield  {journal} {\bibinfo  {journal}
  {Nanoscale}\ }\textbf {\bibinfo {volume} {6}},\ \bibinfo {pages} {3764--3776}
  (\bibinfo {year} {2014})}\BibitemShut {NoStop}%
\bibitem [{\citenamefont {Sakellari}\ \emph {et~al.}(2016)\citenamefont
  {Sakellari}, \citenamefont {Brintakis}, \citenamefont {Kostopoulou},
  \citenamefont {Myrovali}, \citenamefont {Simeonidis}, \citenamefont
  {Lappas},\ and\ \citenamefont {Angelakeris}}]{sakellari2016ferrimagnetic}%
  \BibitemOpen
  \bibfield  {author} {\bibinfo {author} {\bibfnamefont {D.}~\bibnamefont
  {Sakellari}}, \bibinfo {author} {\bibfnamefont {K.}~\bibnamefont
  {Brintakis}}, \bibinfo {author} {\bibfnamefont {A.}~\bibnamefont
  {Kostopoulou}}, \bibinfo {author} {\bibfnamefont {E.}~\bibnamefont
  {Myrovali}}, \bibinfo {author} {\bibfnamefont {K.}~\bibnamefont
  {Simeonidis}}, \bibinfo {author} {\bibfnamefont {A.}~\bibnamefont {Lappas}},
  \ and\ \bibinfo {author} {\bibfnamefont {M.}~\bibnamefont {Angelakeris}},\
  }\bibfield  {title} {\enquote {\bibinfo {title} {Ferrimagnetic nanocrystal
  assemblies as versatile magnetic particle hyperthermia mediators},}\ }\href
  {\doibase 10.1016/j.msec.2015.08.023} {\bibfield  {journal} {\bibinfo
  {journal} {Mater. Sci. Eng., C}\ }\textbf {\bibinfo {volume} {58}},\ \bibinfo
  {pages} {187--193} (\bibinfo {year} {2016})}\BibitemShut {NoStop}%
\bibitem [{\citenamefont {Gavil\'an}\ \emph
  {et~al.}(2017{\natexlab{a}})\citenamefont {Gavil\'an}, \citenamefont
  {Kowalski}, \citenamefont {Heinke}, \citenamefont {Sugunan}, \citenamefont
  {Sommertune}, \citenamefont {Var\'on}, \citenamefont {Bogart}, \citenamefont
  {Posth}, \citenamefont {Zeng}, \citenamefont {Gonz\'alez-Alonso},
  \citenamefont {Balceris}, \citenamefont {Fock}, \citenamefont {Wetterskog},
  \citenamefont {Frandsen}, \citenamefont {Gehrke}, \citenamefont {Gr\"uttner},
  \citenamefont {Fornara}, \citenamefont {Ludwig}, \citenamefont
  {Veintemillas-Verdaguer}, \citenamefont {Johansson},\ and\ \citenamefont
  {Morales}}]{aaa389388fd94f78a5ec927fbdd41d13}%
  \BibitemOpen
  \bibfield  {author} {\bibinfo {author} {\bibfnamefont {H.}~\bibnamefont
  {Gavil\'an}}, \bibinfo {author} {\bibfnamefont {A.}~\bibnamefont {Kowalski}},
  \bibinfo {author} {\bibfnamefont {D.}~\bibnamefont {Heinke}}, \bibinfo
  {author} {\bibfnamefont {A.}~\bibnamefont {Sugunan}}, \bibinfo {author}
  {\bibfnamefont {J.}~\bibnamefont {Sommertune}}, \bibinfo {author}
  {\bibfnamefont {M.}~\bibnamefont {Var\'on}}, \bibinfo {author} {\bibfnamefont
  {L.~K.}\ \bibnamefont {Bogart}}, \bibinfo {author} {\bibfnamefont
  {O.}~\bibnamefont {Posth}}, \bibinfo {author} {\bibfnamefont
  {L.}~\bibnamefont {Zeng}}, \bibinfo {author} {\bibfnamefont {D.}~\bibnamefont
  {Gonz\'alez-Alonso}}, \bibinfo {author} {\bibfnamefont {C.}~\bibnamefont
  {Balceris}}, \bibinfo {author} {\bibfnamefont {J.}~\bibnamefont {Fock}},
  \bibinfo {author} {\bibfnamefont {E.}~\bibnamefont {Wetterskog}}, \bibinfo
  {author} {\bibfnamefont {C.}~\bibnamefont {Frandsen}}, \bibinfo {author}
  {\bibfnamefont {N.}~\bibnamefont {Gehrke}}, \bibinfo {author} {\bibfnamefont
  {C.}~\bibnamefont {Gr\"uttner}}, \bibinfo {author} {\bibfnamefont
  {A.}~\bibnamefont {Fornara}}, \bibinfo {author} {\bibfnamefont
  {F.}~\bibnamefont {Ludwig}}, \bibinfo {author} {\bibfnamefont
  {S.}~\bibnamefont {Veintemillas-Verdaguer}}, \bibinfo {author} {\bibfnamefont
  {C.}~\bibnamefont {Johansson}}, \ and\ \bibinfo {author} {\bibfnamefont
  {M.~P.}\ \bibnamefont {Morales}},\ }\bibfield  {title} {\enquote {\bibinfo
  {title} {Colloidal flower-shaped iron oxide nanoparticles: Synthesis
  strategies and coatings},}\ }\href {\doibase 10.1002/ppsc.201700094}
  {\bibfield  {journal} {\bibinfo  {journal} {Part. Part. Syst. Charact.}\
  }\textbf {\bibinfo {volume} {34}},\ \bibinfo {pages} {1700094} (\bibinfo
  {year} {2017}{\natexlab{a}})}\BibitemShut {NoStop}%
\bibitem [{\citenamefont {Gavil\'an}\ \emph
  {et~al.}(2017{\natexlab{b}})\citenamefont {Gavil\'an}, \citenamefont
  {S\'anchez}, \citenamefont {Brollo}, \citenamefont {As\'in}, \citenamefont
  {Moerner}, \citenamefont {Frandsen}, \citenamefont {L\'azaro}, \citenamefont
  {Serna}, \citenamefont {Veintemillas-Verdaguer}, \citenamefont {Morales},\
  and\ \citenamefont {Guti\'errez}}]{gavilan2017formation}%
  \BibitemOpen
  \bibfield  {author} {\bibinfo {author} {\bibfnamefont {H.}~\bibnamefont
  {Gavil\'an}}, \bibinfo {author} {\bibfnamefont {E.~H.}\ \bibnamefont
  {S\'anchez}}, \bibinfo {author} {\bibfnamefont {M.~E.}\ \bibnamefont
  {Brollo}}, \bibinfo {author} {\bibfnamefont {L.}~\bibnamefont {As\'in}},
  \bibinfo {author} {\bibfnamefont {K.~K.}\ \bibnamefont {Moerner}}, \bibinfo
  {author} {\bibfnamefont {C.}~\bibnamefont {Frandsen}}, \bibinfo {author}
  {\bibfnamefont {F.~J.}\ \bibnamefont {L\'azaro}}, \bibinfo {author}
  {\bibfnamefont {C.~J.}\ \bibnamefont {Serna}}, \bibinfo {author}
  {\bibfnamefont {S.}~\bibnamefont {Veintemillas-Verdaguer}}, \bibinfo {author}
  {\bibfnamefont {M.~P.}\ \bibnamefont {Morales}}, \ and\ \bibinfo {author}
  {\bibfnamefont {L.}~\bibnamefont {Guti\'errez}},\ }\bibfield  {title}
  {\enquote {\bibinfo {title} {Formation mechanism of maghemite nanoflowers
  synthesized by a polyol-mediated process},}\ }\href {\doibase
  10.1021/acsomega.7b00975} {\bibfield  {journal} {\bibinfo  {journal} {ACS
  Omega}\ }\textbf {\bibinfo {volume} {2}},\ \bibinfo {pages} {7172--7184}
  (\bibinfo {year} {2017}{\natexlab{b}})}\BibitemShut {NoStop}%
\bibitem [{\citenamefont {Wetegrove}\ \emph {et~al.}(2019)\citenamefont
  {Wetegrove}, \citenamefont {Witte}, \citenamefont {Bodnar}, \citenamefont
  {Pfahl}, \citenamefont {Springer}, \citenamefont {Schell}, \citenamefont
  {Westphal},\ and\ \citenamefont {Burkel}}]{wetegrove2019formation}%
  \BibitemOpen
  \bibfield  {author} {\bibinfo {author} {\bibfnamefont {M.}~\bibnamefont
  {Wetegrove}}, \bibinfo {author} {\bibfnamefont {K.}~\bibnamefont {Witte}},
  \bibinfo {author} {\bibfnamefont {W.}~\bibnamefont {Bodnar}}, \bibinfo
  {author} {\bibfnamefont {D.-E.}\ \bibnamefont {Pfahl}}, \bibinfo {author}
  {\bibfnamefont {A.}~\bibnamefont {Springer}}, \bibinfo {author}
  {\bibfnamefont {N.}~\bibnamefont {Schell}}, \bibinfo {author} {\bibfnamefont
  {F.}~\bibnamefont {Westphal}}, \ and\ \bibinfo {author} {\bibfnamefont
  {E.}~\bibnamefont {Burkel}},\ }\bibfield  {title} {\enquote {\bibinfo {title}
  {Formation of maghemite nanostructures in polyol: tuning the particle size
  via the precursor stoichiometry},}\ }\href {\doibase 10.1039/C8CE02115E}
  {\bibfield  {journal} {\bibinfo  {journal} {CrystEngComm}\ }\textbf {\bibinfo
  {volume} {21}},\ \bibinfo {pages} {1956--1966} (\bibinfo {year}
  {2019})}\BibitemShut {NoStop}%
\bibitem [{\citenamefont {Shaw}\ \emph {et~al.}(2019)\citenamefont {Shaw},
  \citenamefont {Biswas}, \citenamefont {Gangwar}, \citenamefont {Maiti},
  \citenamefont {Prajapat}, \citenamefont {Meena},\ and\ \citenamefont
  {Prasad}}]{shaw2019synthesis}%
  \BibitemOpen
  \bibfield  {author} {\bibinfo {author} {\bibfnamefont {S.}~\bibnamefont
  {Shaw}}, \bibinfo {author} {\bibfnamefont {A.}~\bibnamefont {Biswas}},
  \bibinfo {author} {\bibfnamefont {A.}~\bibnamefont {Gangwar}}, \bibinfo
  {author} {\bibfnamefont {P.}~\bibnamefont {Maiti}}, \bibinfo {author}
  {\bibfnamefont {C.}~\bibnamefont {Prajapat}}, \bibinfo {author}
  {\bibfnamefont {S.~S.}\ \bibnamefont {Meena}}, \ and\ \bibinfo {author}
  {\bibfnamefont {N.}~\bibnamefont {Prasad}},\ }\bibfield  {title} {\enquote
  {\bibinfo {title} {Synthesis of exchange coupled nanoflowers for efficient
  magnetic hyperthermia},}\ }\href {\doibase 10.1016/j.jmmm.2019.04.056}
  {\bibfield  {journal} {\bibinfo  {journal} {J. Magn. Magn. Mater.}\ }\textbf
  {\bibinfo {volume} {484}},\ \bibinfo {pages} {437--444} (\bibinfo {year}
  {2019})}\BibitemShut {NoStop}%
\bibitem [{\citenamefont {Alonso}\ \emph {et~al.}(2010)\citenamefont {Alonso},
  \citenamefont {Fdez-Gubieda}, \citenamefont {Barandiar{\'a}n}, \citenamefont
  {Svalov}, \citenamefont {Barqu{\'\i}n}, \citenamefont {Venero},\ and\
  \citenamefont {Orue}}]{alonso2010crossover}%
  \BibitemOpen
  \bibfield  {author} {\bibinfo {author} {\bibfnamefont {J.}~\bibnamefont
  {Alonso}}, \bibinfo {author} {\bibfnamefont {M.}~\bibnamefont
  {Fdez-Gubieda}}, \bibinfo {author} {\bibfnamefont {J.}~\bibnamefont
  {Barandiar{\'a}n}}, \bibinfo {author} {\bibfnamefont {A.}~\bibnamefont
  {Svalov}}, \bibinfo {author} {\bibfnamefont {L.~F.}\ \bibnamefont
  {Barqu{\'\i}n}}, \bibinfo {author} {\bibfnamefont {D.~A.}\ \bibnamefont
  {Venero}}, \ and\ \bibinfo {author} {\bibfnamefont {I.}~\bibnamefont
  {Orue}},\ }\bibfield  {title} {\enquote {\bibinfo {title} {Crossover from
  superspin glass to superferromagnet in $\mathrm{Fe_xAg_{100- x}}$
  nanostructured thin films ($20\leq x \leq 50$)},}\ }\href {\doibase
  10.1103/PhysRevB.82.054406} {\bibfield  {journal} {\bibinfo  {journal} {Phys.
  Rev. B}\ }\textbf {\bibinfo {volume} {82}},\ \bibinfo {pages} {054406}
  (\bibinfo {year} {2010})}\BibitemShut {NoStop}%
\bibitem [{\citenamefont {Dutz}(2016)}]{dutz2016magnetic}%
  \BibitemOpen
  \bibfield  {author} {\bibinfo {author} {\bibfnamefont {S.}~\bibnamefont
  {Dutz}},\ }\bibfield  {title} {\enquote {\bibinfo {title} {Are magnetic
  multicore nanoparticles promising candidates for biomedical applications?}}\
  }\href {\doibase 10.1109/TMAG.2016.2570745} {\bibfield  {journal} {\bibinfo
  {journal} {IEEE Trans. Magn.}\ }\textbf {\bibinfo {volume} {52}},\ \bibinfo
  {pages} {1--3} (\bibinfo {year} {2016})}\BibitemShut {NoStop}%
\bibitem [{\citenamefont {D{\"o}brich}\ \emph {et~al.}(2009)\citenamefont
  {D{\"o}brich}, \citenamefont {Elmas}, \citenamefont {Ferdinand},
  \citenamefont {Markmann}, \citenamefont {Sharp}, \citenamefont {Eckerlebe},
  \citenamefont {Kohlbrecher}, \citenamefont {Birringer},\ and\ \citenamefont
  {Michels}}]{dobrich2009grain}%
  \BibitemOpen
  \bibfield  {author} {\bibinfo {author} {\bibfnamefont {F.}~\bibnamefont
  {D{\"o}brich}}, \bibinfo {author} {\bibfnamefont {M.}~\bibnamefont {Elmas}},
  \bibinfo {author} {\bibfnamefont {A.}~\bibnamefont {Ferdinand}}, \bibinfo
  {author} {\bibfnamefont {J.}~\bibnamefont {Markmann}}, \bibinfo {author}
  {\bibfnamefont {M.}~\bibnamefont {Sharp}}, \bibinfo {author} {\bibfnamefont
  {H.}~\bibnamefont {Eckerlebe}}, \bibinfo {author} {\bibfnamefont
  {J.}~\bibnamefont {Kohlbrecher}}, \bibinfo {author} {\bibfnamefont
  {R.}~\bibnamefont {Birringer}}, \ and\ \bibinfo {author} {\bibfnamefont
  {A.}~\bibnamefont {Michels}},\ }\bibfield  {title} {\enquote {\bibinfo
  {title} {Grain-boundary-induced spin disorder in nanocrystalline
  gadolinium},}\ }\href {\doibase 10.1088/0953-8984/21/15/156003} {\bibfield
  {journal} {\bibinfo  {journal} {J. Phys.: Condens. Matter}\ }\textbf
  {\bibinfo {volume} {21}},\ \bibinfo {pages} {156003} (\bibinfo {year}
  {2009})}\BibitemShut {NoStop}%
\bibitem [{\citenamefont {Bender}\ \emph
  {et~al.}(2018{\natexlab{a}})\citenamefont {Bender}, \citenamefont {Fock},
  \citenamefont {Frandsen}, \citenamefont {Hansen}, \citenamefont {Balceris},
  \citenamefont {Ludwig}, \citenamefont {Posth}, \citenamefont {Wetterskog},
  \citenamefont {Bogart}, \citenamefont {Southern}, \citenamefont {Szczerba},
  \citenamefont {Zeng}, \citenamefont {Witte}, \citenamefont {Gr\"uttner},
  \citenamefont {Westphal}, \citenamefont {Honecker}, \citenamefont
  {Gonz\'alez-Alonso}, \citenamefont {Fern\'andez~Barqu\'in},\ and\
  \citenamefont {Johansson}}]{bender2018relating}%
  \BibitemOpen
  \bibfield  {author} {\bibinfo {author} {\bibfnamefont {P.}~\bibnamefont
  {Bender}}, \bibinfo {author} {\bibfnamefont {J.}~\bibnamefont {Fock}},
  \bibinfo {author} {\bibfnamefont {C.}~\bibnamefont {Frandsen}}, \bibinfo
  {author} {\bibfnamefont {M.~F.}\ \bibnamefont {Hansen}}, \bibinfo {author}
  {\bibfnamefont {C.}~\bibnamefont {Balceris}}, \bibinfo {author}
  {\bibfnamefont {F.}~\bibnamefont {Ludwig}}, \bibinfo {author} {\bibfnamefont
  {O.}~\bibnamefont {Posth}}, \bibinfo {author} {\bibfnamefont
  {E.}~\bibnamefont {Wetterskog}}, \bibinfo {author} {\bibfnamefont {L.~K.}\
  \bibnamefont {Bogart}}, \bibinfo {author} {\bibfnamefont {P.}~\bibnamefont
  {Southern}}, \bibinfo {author} {\bibfnamefont {W.}~\bibnamefont {Szczerba}},
  \bibinfo {author} {\bibfnamefont {L.}~\bibnamefont {Zeng}}, \bibinfo {author}
  {\bibfnamefont {K.}~\bibnamefont {Witte}}, \bibinfo {author} {\bibfnamefont
  {C.}~\bibnamefont {Gr\"uttner}}, \bibinfo {author} {\bibfnamefont
  {F.}~\bibnamefont {Westphal}}, \bibinfo {author} {\bibfnamefont
  {D.}~\bibnamefont {Honecker}}, \bibinfo {author} {\bibfnamefont
  {D.}~\bibnamefont {Gonz\'alez-Alonso}}, \bibinfo {author} {\bibfnamefont
  {L.}~\bibnamefont {Fern\'andez~Barqu\'in}}, \ and\ \bibinfo {author}
  {\bibfnamefont {C.}~\bibnamefont {Johansson}},\ }\bibfield  {title} {\enquote
  {\bibinfo {title} {Relating magnetic properties and high hyperthermia
  performance of iron oxide nanoflowers},}\ }\href {\doibase
  10.1021/acs.jpcc.7b11255} {\bibfield  {journal} {\bibinfo  {journal} {J.
  Phys. Chem. C}\ }\textbf {\bibinfo {volume} {122}},\ \bibinfo {pages}
  {3068--3077} (\bibinfo {year} {2018}{\natexlab{a}})}\BibitemShut {NoStop}%
\bibitem [{\citenamefont {Bender}\ \emph
  {et~al.}(2018{\natexlab{b}})\citenamefont {Bender}, \citenamefont {Fock},
  \citenamefont {Hansen}, \citenamefont {Bogart}, \citenamefont {Southern},
  \citenamefont {Ludwig}, \citenamefont {Wiekhorst}, \citenamefont {Szczerba},
  \citenamefont {Zeng}, \citenamefont {Heinke}, \citenamefont {Gehrke},
  \citenamefont {Fern\'andez~D\'iaz}, \citenamefont {Gonz\'alez-Alonso},
  \citenamefont {Espeso}, \citenamefont {Rodr\'iguez~Fern\'andez},\ and\
  \citenamefont {Johansson}}]{bender2018influence}%
  \BibitemOpen
  \bibfield  {author} {\bibinfo {author} {\bibfnamefont {P.}~\bibnamefont
  {Bender}}, \bibinfo {author} {\bibfnamefont {J.}~\bibnamefont {Fock}},
  \bibinfo {author} {\bibfnamefont {M.}~\bibnamefont {Hansen}}, \bibinfo
  {author} {\bibfnamefont {L.}~\bibnamefont {Bogart}}, \bibinfo {author}
  {\bibfnamefont {P.}~\bibnamefont {Southern}}, \bibinfo {author}
  {\bibfnamefont {F.}~\bibnamefont {Ludwig}}, \bibinfo {author} {\bibfnamefont
  {F.}~\bibnamefont {Wiekhorst}}, \bibinfo {author} {\bibfnamefont
  {W.}~\bibnamefont {Szczerba}}, \bibinfo {author} {\bibfnamefont
  {L.}~\bibnamefont {Zeng}}, \bibinfo {author} {\bibfnamefont {D.}~\bibnamefont
  {Heinke}}, \bibinfo {author} {\bibfnamefont {N.}~\bibnamefont {Gehrke}},
  \bibinfo {author} {\bibfnamefont {M.~T.}\ \bibnamefont {Fern\'andez~D\'iaz}},
  \bibinfo {author} {\bibfnamefont {D.}~\bibnamefont {Gonz\'alez-Alonso}},
  \bibinfo {author} {\bibfnamefont {J.~I.}\ \bibnamefont {Espeso}}, \bibinfo
  {author} {\bibfnamefont {J.}~\bibnamefont {Rodr\'iguez~Fern\'andez}}, \ and\
  \bibinfo {author} {\bibfnamefont {C.}~\bibnamefont {Johansson}},\ }\bibfield
  {title} {\enquote {\bibinfo {title} {Influence of clustering on the magnetic
  properties and hyperthermia performance of iron oxide nanoparticles},}\
  }\href {\doibase 10.1088/1361-6528/aad67d} {\bibfield  {journal} {\bibinfo
  {journal} {Nanotechnology}\ }\textbf {\bibinfo {volume} {29}},\ \bibinfo
  {pages} {425705} (\bibinfo {year} {2018}{\natexlab{b}})}\BibitemShut
  {NoStop}%
\bibitem [{\citenamefont {Lak}\ \emph {et~al.}(2018)\citenamefont {Lak},
  \citenamefont {Cassani}, \citenamefont {Mai}, \citenamefont {Winckelmans},
  \citenamefont {Cabrera}, \citenamefont {Sadrollahi}, \citenamefont {Marras},
  \citenamefont {Remmer}, \citenamefont {Fiorito}, \citenamefont
  {Cremades-Jimeno}, \citenamefont {Litterst}, \citenamefont {Ludwig},
  \citenamefont {Liberato}, \citenamefont {Teran}, \citenamefont {Bals},\ and\
  \citenamefont {Pellegrino}}]{lak2018fe2+}%
  \BibitemOpen
  \bibfield  {author} {\bibinfo {author} {\bibfnamefont {A.}~\bibnamefont
  {Lak}}, \bibinfo {author} {\bibfnamefont {M.}~\bibnamefont {Cassani}},
  \bibinfo {author} {\bibfnamefont {B.~T.}\ \bibnamefont {Mai}}, \bibinfo
  {author} {\bibfnamefont {N.}~\bibnamefont {Winckelmans}}, \bibinfo {author}
  {\bibfnamefont {D.}~\bibnamefont {Cabrera}}, \bibinfo {author} {\bibfnamefont
  {E.}~\bibnamefont {Sadrollahi}}, \bibinfo {author} {\bibfnamefont
  {S.}~\bibnamefont {Marras}}, \bibinfo {author} {\bibfnamefont
  {H.}~\bibnamefont {Remmer}}, \bibinfo {author} {\bibfnamefont
  {S.}~\bibnamefont {Fiorito}}, \bibinfo {author} {\bibfnamefont
  {L.}~\bibnamefont {Cremades-Jimeno}}, \bibinfo {author} {\bibfnamefont
  {F.~J.}\ \bibnamefont {Litterst}}, \bibinfo {author} {\bibfnamefont
  {F.}~\bibnamefont {Ludwig}}, \bibinfo {author} {\bibfnamefont
  {M.}~\bibnamefont {Liberato}}, \bibinfo {author} {\bibfnamefont {F.~J.}\
  \bibnamefont {Teran}}, \bibinfo {author} {\bibfnamefont {S.}~\bibnamefont
  {Bals}}, \ and\ \bibinfo {author} {\bibfnamefont {T.}~\bibnamefont
  {Pellegrino}},\ }\bibfield  {title} {\enquote {\bibinfo {title}
  {$\mathrm{Fe^{2+}}$ deficiencies, $\mathrm{FeO}$ subdomains, and structural
  defects favor magnetic hyperthermia performance of iron oxide nanocubes into
  intracellular environment},}\ }\href {\doibase 10.1021/acs.nanolett.8b02722}
  {\bibfield  {journal} {\bibinfo  {journal} {Nano Lett.}\ }\textbf {\bibinfo
  {volume} {18}},\ \bibinfo {pages} {6856--6866} (\bibinfo {year}
  {2018})}\BibitemShut {NoStop}%
\bibitem [{\citenamefont {Kuznetsov}(2019)}]{kuznetsov2019zero}%
  \BibitemOpen
  \bibfield  {author} {\bibinfo {author} {\bibfnamefont {A.~A.}\ \bibnamefont
  {Kuznetsov}},\ }\bibfield  {title} {\enquote {\bibinfo {title} {Zero-field
  and field-induced interactions between multicore magnetic nanoparticles},}\
  }\href {\doibase 10.3390/nano9050718} {\bibfield  {journal} {\bibinfo
  {journal} {Nanomaterials}\ }\textbf {\bibinfo {volume} {9}},\ \bibinfo
  {pages} {718} (\bibinfo {year} {2019})}\BibitemShut {NoStop}%
\bibitem [{\citenamefont {Bender}\ \emph
  {et~al.}(2018{\natexlab{c}})\citenamefont {Bender}, \citenamefont
  {Wetterskog}, \citenamefont {Honecker}, \citenamefont {Fock}, \citenamefont
  {Frandsen}, \citenamefont {Moerland}, \citenamefont {Bogart}, \citenamefont
  {Posth}, \citenamefont {Szczerba}, \citenamefont {Gavil{\'a}n}, \citenamefont
  {Costo}, \citenamefont {Fern\'andez-D\'iaz}, \citenamefont
  {Gonz\'alez-Alonso}, \citenamefont {Fern\'andez~Barqu\'in},\ and\
  \citenamefont {Johansson}}]{bender2018dipolar}%
  \BibitemOpen
  \bibfield  {author} {\bibinfo {author} {\bibfnamefont {P.}~\bibnamefont
  {Bender}}, \bibinfo {author} {\bibfnamefont {E.}~\bibnamefont {Wetterskog}},
  \bibinfo {author} {\bibfnamefont {D.}~\bibnamefont {Honecker}}, \bibinfo
  {author} {\bibfnamefont {J.}~\bibnamefont {Fock}}, \bibinfo {author}
  {\bibfnamefont {C.}~\bibnamefont {Frandsen}}, \bibinfo {author}
  {\bibfnamefont {C.}~\bibnamefont {Moerland}}, \bibinfo {author}
  {\bibfnamefont {L.~K.}\ \bibnamefont {Bogart}}, \bibinfo {author}
  {\bibfnamefont {O.}~\bibnamefont {Posth}}, \bibinfo {author} {\bibfnamefont
  {W.}~\bibnamefont {Szczerba}}, \bibinfo {author} {\bibfnamefont
  {H.}~\bibnamefont {Gavil{\'a}n}}, \bibinfo {author} {\bibfnamefont
  {R.}~\bibnamefont {Costo}}, \bibinfo {author} {\bibfnamefont {M.~T.}\
  \bibnamefont {Fern\'andez-D\'iaz}}, \bibinfo {author} {\bibfnamefont
  {D.}~\bibnamefont {Gonz\'alez-Alonso}}, \bibinfo {author} {\bibfnamefont
  {L.}~\bibnamefont {Fern\'andez~Barqu\'in}}, \ and\ \bibinfo {author}
  {\bibfnamefont {C.}~\bibnamefont {Johansson}},\ }\bibfield  {title} {\enquote
  {\bibinfo {title} {Dipolar-coupled moment correlations in clusters of
  magnetic nanoparticles},}\ }\href {\doibase 10.1103/PhysRevB.98.224420}
  {\bibfield  {journal} {\bibinfo  {journal} {Phys. Rev. B}\ }\textbf {\bibinfo
  {volume} {98}},\ \bibinfo {pages} {224420} (\bibinfo {year}
  {2018}{\natexlab{c}})}\BibitemShut {NoStop}%
\bibitem [{\citenamefont {Bender}\ \emph
  {et~al.}(2017{\natexlab{a}})\citenamefont {Bender}, \citenamefont
  {Fern\'{a}ndez~Barqu\'{i}n}, \citenamefont {Fdez-Gubieda}, \citenamefont
  {Gonz\'{a}lez-Alonso}, \citenamefont {Honecker}, \citenamefont {Marcano},\
  and\ \citenamefont {Szczerba}}]{ILLproposal}%
  \BibitemOpen
  \bibfield  {author} {\bibinfo {author} {\bibfnamefont {P.}~\bibnamefont
  {Bender}}, \bibinfo {author} {\bibfnamefont {L.}~\bibnamefont
  {Fern\'{a}ndez~Barqu\'{i}n}}, \bibinfo {author} {\bibfnamefont {M.-L.}\
  \bibnamefont {Fdez-Gubieda}}, \bibinfo {author} {\bibfnamefont
  {D.}~\bibnamefont {Gonz\'{a}lez-Alonso}}, \bibinfo {author} {\bibfnamefont
  {D.}~\bibnamefont {Honecker}}, \bibinfo {author} {\bibfnamefont
  {L.}~\bibnamefont {Marcano}}, \ and\ \bibinfo {author} {\bibfnamefont
  {W.}~\bibnamefont {Szczerba}},\ }\bibfield  {title} {\enquote {\bibinfo
  {title} {Spin correlation in clusters of magnetic nanoparticles.}}\ }\href
  {\doibase 10.5291/ILL-DATA.5-53-267} {\bibfield  {journal} {\bibinfo
  {journal} {Institut Laue-Langevin (ILL)}\ } (\bibinfo {year}
  {2017}{\natexlab{a}}),\ 10.5291/ILL-DATA.5-53-267}\BibitemShut {NoStop}%
\bibitem [{\citenamefont {Dewhurst}\ \emph {et~al.}(2016)\citenamefont
  {Dewhurst}, \citenamefont {Grillo}, \citenamefont {Honecker}, \citenamefont
  {Bonnaud}, \citenamefont {Jacques}, \citenamefont {Amrouni}, \citenamefont
  {Perillo-Marcone}, \citenamefont {Manzin},\ and\ \citenamefont
  {Cubitt}}]{Dewhurst:ks5488}%
  \BibitemOpen
  \bibfield  {author} {\bibinfo {author} {\bibfnamefont {C.~D.}\ \bibnamefont
  {Dewhurst}}, \bibinfo {author} {\bibfnamefont {I.}~\bibnamefont {Grillo}},
  \bibinfo {author} {\bibfnamefont {D.}~\bibnamefont {Honecker}}, \bibinfo
  {author} {\bibfnamefont {M.}~\bibnamefont {Bonnaud}}, \bibinfo {author}
  {\bibfnamefont {M.}~\bibnamefont {Jacques}}, \bibinfo {author} {\bibfnamefont
  {C.}~\bibnamefont {Amrouni}}, \bibinfo {author} {\bibfnamefont
  {A.}~\bibnamefont {Perillo-Marcone}}, \bibinfo {author} {\bibfnamefont
  {G.}~\bibnamefont {Manzin}}, \ and\ \bibinfo {author} {\bibfnamefont
  {R.}~\bibnamefont {Cubitt}},\ }\bibfield  {title} {\enquote {\bibinfo {title}
  {{The small-angle neutron scattering instrument D33 at the Institut
  Laue{--}Langevin}},}\ }\href {\doibase 10.1107/S1600576715021792} {\bibfield
  {journal} {\bibinfo  {journal} {J. Appl. Crystallogr.}\ }\textbf {\bibinfo
  {volume} {49}},\ \bibinfo {pages} {1--14} (\bibinfo {year}
  {2016})}\BibitemShut {NoStop}%
\bibitem [{\citenamefont {Krycka}\ \emph {et~al.}(2010)\citenamefont {Krycka},
  \citenamefont {Booth}, \citenamefont {Hogg}, \citenamefont {Ijiri},
  \citenamefont {Borchers}, \citenamefont {Chen}, \citenamefont {Watson},
  \citenamefont {Laver}, \citenamefont {Gentile}, \citenamefont {Dedon},
  \citenamefont {Harris}, \citenamefont {Rhyne},\ and\ \citenamefont
  {Majetich}}]{krycka2010core}%
  \BibitemOpen
  \bibfield  {author} {\bibinfo {author} {\bibfnamefont {K.~L.}\ \bibnamefont
  {Krycka}}, \bibinfo {author} {\bibfnamefont {R.~A.}\ \bibnamefont {Booth}},
  \bibinfo {author} {\bibfnamefont {C.~R.}\ \bibnamefont {Hogg}}, \bibinfo
  {author} {\bibfnamefont {Y.}~\bibnamefont {Ijiri}}, \bibinfo {author}
  {\bibfnamefont {J.~A.}\ \bibnamefont {Borchers}}, \bibinfo {author}
  {\bibfnamefont {W.}~\bibnamefont {Chen}}, \bibinfo {author} {\bibfnamefont
  {S.~M.}\ \bibnamefont {Watson}}, \bibinfo {author} {\bibfnamefont
  {M.}~\bibnamefont {Laver}}, \bibinfo {author} {\bibfnamefont {T.~R.}\
  \bibnamefont {Gentile}}, \bibinfo {author} {\bibfnamefont {L.~R.}\
  \bibnamefont {Dedon}}, \bibinfo {author} {\bibfnamefont {S.}~\bibnamefont
  {Harris}}, \bibinfo {author} {\bibfnamefont {J.~J.}\ \bibnamefont {Rhyne}}, \
  and\ \bibinfo {author} {\bibfnamefont {S.~A.}\ \bibnamefont {Majetich}},\
  }\bibfield  {title} {\enquote {\bibinfo {title} {Core-shell magnetic
  morphology of structurally uniform magnetite nanoparticles},}\ }\href
  {\doibase 10.1103/PhysRevLett.104.207203} {\bibfield  {journal} {\bibinfo
  {journal} {Phys. Rev. Lett.}\ }\textbf {\bibinfo {volume} {104}},\ \bibinfo
  {pages} {207203} (\bibinfo {year} {2010})}\BibitemShut {NoStop}%
\bibitem [{\citenamefont {Orue}\ \emph {et~al.}(2018)\citenamefont {Orue},
  \citenamefont {Marcano}, \citenamefont {Bender}, \citenamefont
  {Garc\'{i}a-Prieto}, \citenamefont {Valencia}, \citenamefont {Mawass},
  \citenamefont {Gil-Cart\'{o}n}, \citenamefont {Alba~Venero}, \citenamefont
  {Honecker}, \citenamefont {Garc\'{i}a-Arribas}, \citenamefont
  {Fern\'{a}ndez~Barqu\'{i}n}, \citenamefont {Muela},\ and\ \citenamefont
  {Fdez-Gubieda}}]{Orue2018}%
  \BibitemOpen
  \bibfield  {author} {\bibinfo {author} {\bibfnamefont {I.}~\bibnamefont
  {Orue}}, \bibinfo {author} {\bibfnamefont {L.}~\bibnamefont {Marcano}},
  \bibinfo {author} {\bibfnamefont {P.}~\bibnamefont {Bender}}, \bibinfo
  {author} {\bibfnamefont {A.}~\bibnamefont {Garc\'{i}a-Prieto}}, \bibinfo
  {author} {\bibfnamefont {S.}~\bibnamefont {Valencia}}, \bibinfo {author}
  {\bibfnamefont {M.~A.}\ \bibnamefont {Mawass}}, \bibinfo {author}
  {\bibfnamefont {D.}~\bibnamefont {Gil-Cart\'{o}n}}, \bibinfo {author}
  {\bibfnamefont {D.}~\bibnamefont {Alba~Venero}}, \bibinfo {author}
  {\bibfnamefont {D.}~\bibnamefont {Honecker}}, \bibinfo {author}
  {\bibfnamefont {A.}~\bibnamefont {Garc\'{i}a-Arribas}}, \bibinfo {author}
  {\bibfnamefont {L.}~\bibnamefont {Fern\'{a}ndez~Barqu\'{i}n}}, \bibinfo
  {author} {\bibfnamefont {A.}~\bibnamefont {Muela}}, \ and\ \bibinfo {author}
  {\bibfnamefont {M.~L.}\ \bibnamefont {Fdez-Gubieda}},\ }\bibfield  {title}
  {\enquote {\bibinfo {title} {Configuration of the magnetosome chain: a
  natural magnetic nanoarchitecture},}\ }\href {\doibase 10.1039/C7NR08493E}
  {\bibfield  {journal} {\bibinfo  {journal} {Nanoscale}\ }\textbf {\bibinfo
  {volume} {10}},\ \bibinfo {pages} {7407--7419} (\bibinfo {year}
  {2018})}\BibitemShut {NoStop}%
\bibitem [{\citenamefont {M{\"u}hlbauer}\ \emph {et~al.}(2019)\citenamefont
  {M{\"u}hlbauer}, \citenamefont {Honecker}, \citenamefont {P{\'e}rigo},
  \citenamefont {Bergner}, \citenamefont {Disch}, \citenamefont {Heinemann},
  \citenamefont {Erokhin}, \citenamefont {Berkov}, \citenamefont {Leighton},
  \citenamefont {Eskildsen},\ and\ \citenamefont
  {Michels}}]{muhlbauer2019magnetic}%
  \BibitemOpen
  \bibfield  {author} {\bibinfo {author} {\bibfnamefont {S.}~\bibnamefont
  {M{\"u}hlbauer}}, \bibinfo {author} {\bibfnamefont {D.}~\bibnamefont
  {Honecker}}, \bibinfo {author} {\bibfnamefont {{\'E}.~A.}\ \bibnamefont
  {P{\'e}rigo}}, \bibinfo {author} {\bibfnamefont {F.}~\bibnamefont {Bergner}},
  \bibinfo {author} {\bibfnamefont {S.}~\bibnamefont {Disch}}, \bibinfo
  {author} {\bibfnamefont {A.}~\bibnamefont {Heinemann}}, \bibinfo {author}
  {\bibfnamefont {S.}~\bibnamefont {Erokhin}}, \bibinfo {author} {\bibfnamefont
  {D.}~\bibnamefont {Berkov}}, \bibinfo {author} {\bibfnamefont
  {C.}~\bibnamefont {Leighton}}, \bibinfo {author} {\bibfnamefont {M.~R.}\
  \bibnamefont {Eskildsen}}, \ and\ \bibinfo {author} {\bibfnamefont
  {A.}~\bibnamefont {Michels}},\ }\bibfield  {title} {\enquote {\bibinfo
  {title} {Magnetic small-angle neutron scattering},}\ }\href {\doibase
  10.1103/RevModPhys.91.015004} {\bibfield  {journal} {\bibinfo  {journal}
  {Rev. Mod. Phys.}\ }\textbf {\bibinfo {volume} {91}},\ \bibinfo {pages}
  {015004} (\bibinfo {year} {2019})}\BibitemShut {NoStop}%
\bibitem [{\citenamefont {Honecker}\ \emph {et~al.}(2010)\citenamefont
  {Honecker}, \citenamefont {Ferdinand}, \citenamefont {D{\"o}brich},
  \citenamefont {Dewhurst}, \citenamefont {Wiedenmann}, \citenamefont
  {G{\'o}mez-Polo}, \citenamefont {Suzuki},\ and\ \citenamefont
  {Michels}}]{honecker2010longitudinal}%
  \BibitemOpen
  \bibfield  {author} {\bibinfo {author} {\bibfnamefont {D.}~\bibnamefont
  {Honecker}}, \bibinfo {author} {\bibfnamefont {A.}~\bibnamefont {Ferdinand}},
  \bibinfo {author} {\bibfnamefont {F.}~\bibnamefont {D{\"o}brich}}, \bibinfo
  {author} {\bibfnamefont {C.~D.}\ \bibnamefont {Dewhurst}}, \bibinfo {author}
  {\bibfnamefont {A.}~\bibnamefont {Wiedenmann}}, \bibinfo {author}
  {\bibfnamefont {C.}~\bibnamefont {G{\'o}mez-Polo}}, \bibinfo {author}
  {\bibfnamefont {K.}~\bibnamefont {Suzuki}}, \ and\ \bibinfo {author}
  {\bibfnamefont {A.}~\bibnamefont {Michels}},\ }\bibfield  {title} {\enquote
  {\bibinfo {title} {Longitudinal polarization analysis in small-angle neutron
  scattering},}\ }\href {\doibase 10.1140/epjb/e2010-00191-5} {\bibfield
  {journal} {\bibinfo  {journal} {Eur. Phys. J. B}\ }\textbf {\bibinfo {volume}
  {76}},\ \bibinfo {pages} {209--213} (\bibinfo {year} {2010})}\BibitemShut
  {NoStop}%
\bibitem [{\citenamefont {Pedersen}(1997)}]{pedersen1997}%
  \BibitemOpen
  \bibfield  {author} {\bibinfo {author} {\bibfnamefont {J.~S.}\ \bibnamefont
  {Pedersen}},\ }\bibfield  {title} {\enquote {\bibinfo {title} {Analysis of
  small-angle scattering data from colloids and polymer solutions: modeling and
  least-squares fitting},}\ }\href {\doibase 10.1016/S0001-8686(97)00312-6}
  {\bibfield  {journal} {\bibinfo  {journal} {Adv. Colloid Interface Sci.}\
  }\textbf {\bibinfo {volume} {70}},\ \bibinfo {pages} {171 -- 210} (\bibinfo
  {year} {1997})}\BibitemShut {NoStop}%
\bibitem [{\citenamefont {Li}, \citenamefont {Senesi},\ and\ \citenamefont
  {Lee}(2016)}]{li2016small}%
  \BibitemOpen
  \bibfield  {author} {\bibinfo {author} {\bibfnamefont {T.}~\bibnamefont
  {Li}}, \bibinfo {author} {\bibfnamefont {A.~J.}\ \bibnamefont {Senesi}}, \
  and\ \bibinfo {author} {\bibfnamefont {B.}~\bibnamefont {Lee}},\ }\bibfield
  {title} {\enquote {\bibinfo {title} {Small angle x-ray scattering for
  nanoparticle research},}\ }\href {\doibase 10.1021/acs.chemrev.5b00690}
  {\bibfield  {journal} {\bibinfo  {journal} {Chem. Rev.}\ }\textbf {\bibinfo
  {volume} {116}},\ \bibinfo {pages} {11128--11180} (\bibinfo {year}
  {2016})}\BibitemShut {NoStop}%
\bibitem [{\citenamefont {Alba~Venero}\ \emph {et~al.}(2016)\citenamefont
  {Alba~Venero}, \citenamefont {Rogers}, \citenamefont {Langridge},
  \citenamefont {Alonso}, \citenamefont {Fdez-Gubieda}, \citenamefont
  {Svalov},\ and\ \citenamefont
  {Fern{\'a}ndez~Barqu{\'\i}n}}]{alba2016magnetic}%
  \BibitemOpen
  \bibfield  {author} {\bibinfo {author} {\bibfnamefont {D.}~\bibnamefont
  {Alba~Venero}}, \bibinfo {author} {\bibfnamefont {S.}~\bibnamefont {Rogers}},
  \bibinfo {author} {\bibfnamefont {S.}~\bibnamefont {Langridge}}, \bibinfo
  {author} {\bibfnamefont {J.}~\bibnamefont {Alonso}}, \bibinfo {author}
  {\bibfnamefont {M.}~\bibnamefont {Fdez-Gubieda}}, \bibinfo {author}
  {\bibfnamefont {A.}~\bibnamefont {Svalov}}, \ and\ \bibinfo {author}
  {\bibfnamefont {L.}~\bibnamefont {Fern{\'a}ndez~Barqu{\'\i}n}},\ }\bibfield
  {title} {\enquote {\bibinfo {title} {Magnetic nanoscopic correlations in the
  crossover between a superspin glass and a superferromagnet},}\ }\href
  {\doibase 10.1063/1.4945427} {\bibfield  {journal} {\bibinfo  {journal} {J.
  Appl. Phys.}\ }\textbf {\bibinfo {volume} {119}},\ \bibinfo {pages} {143902}
  (\bibinfo {year} {2016})}\BibitemShut {NoStop}%
\bibitem [{\citenamefont {Honecker}, \citenamefont {Barqu{\'\i}n},\ and\
  \citenamefont {Bender}(2019)}]{honecker2019magnetic}%
  \BibitemOpen
  \bibfield  {author} {\bibinfo {author} {\bibfnamefont {D.}~\bibnamefont
  {Honecker}}, \bibinfo {author} {\bibfnamefont {L.~F.}\ \bibnamefont
  {Barqu{\'\i}n}}, \ and\ \bibinfo {author} {\bibfnamefont {P.}~\bibnamefont
  {Bender}},\ }\bibfield  {title} {\enquote {\bibinfo {title} {The magnetic
  structure factor of correlated nanoparticle moments in small-angle neutron
  scattering},}\ }\href {https://arxiv.org/abs/1904.06243} {\bibfield
  {journal} {\bibinfo  {journal} {Preprint at
  https://arxiv.org/abs/1904.06243}\ } (\bibinfo {year} {2019})}\BibitemShut
  {NoStop}%
\bibitem [{\citenamefont {Bender}\ \emph
  {et~al.}(2017{\natexlab{b}})\citenamefont {Bender}, \citenamefont {Bogart},
  \citenamefont {Posth}, \citenamefont {Szczerba}, \citenamefont {Rogers},
  \citenamefont {Castro}, \citenamefont {Nilsson}, \citenamefont {Zeng},
  \citenamefont {Sugunan}, \citenamefont {Sommertune}, \citenamefont {Fornara},
  \citenamefont {Gonz\'alez-Alonso}, \citenamefont {Fern\'andez~Barqu\'in},\
  and\ \citenamefont {Johansson}}]{bender2017structural}%
  \BibitemOpen
  \bibfield  {author} {\bibinfo {author} {\bibfnamefont {P.}~\bibnamefont
  {Bender}}, \bibinfo {author} {\bibfnamefont {L.}~\bibnamefont {Bogart}},
  \bibinfo {author} {\bibfnamefont {O.}~\bibnamefont {Posth}}, \bibinfo
  {author} {\bibfnamefont {W.}~\bibnamefont {Szczerba}}, \bibinfo {author}
  {\bibfnamefont {S.}~\bibnamefont {Rogers}}, \bibinfo {author} {\bibfnamefont
  {A.}~\bibnamefont {Castro}}, \bibinfo {author} {\bibfnamefont
  {L.}~\bibnamefont {Nilsson}}, \bibinfo {author} {\bibfnamefont
  {L.}~\bibnamefont {Zeng}}, \bibinfo {author} {\bibfnamefont {A.}~\bibnamefont
  {Sugunan}}, \bibinfo {author} {\bibfnamefont {J.}~\bibnamefont {Sommertune}},
  \bibinfo {author} {\bibfnamefont {A.}~\bibnamefont {Fornara}}, \bibinfo
  {author} {\bibfnamefont {D.}~\bibnamefont {Gonz\'alez-Alonso}}, \bibinfo
  {author} {\bibfnamefont {L.}~\bibnamefont {Fern\'andez~Barqu\'in}}, \ and\
  \bibinfo {author} {\bibfnamefont {C.}~\bibnamefont {Johansson}},\ }\bibfield
  {title} {\enquote {\bibinfo {title} {Structural and magnetic properties of
  multi-core nanoparticles analysed using a generalised numerical inversion
  method},}\ }\href {\doibase 10.1038/srep45990} {\bibfield  {journal}
  {\bibinfo  {journal} {Sci. Rep.}\ }\textbf {\bibinfo {volume} {7}},\ \bibinfo
  {pages} {45990} (\bibinfo {year} {2017}{\natexlab{b}})}\BibitemShut {NoStop}%
\bibitem [{\citenamefont {Urtizberea}\ \emph {et~al.}(2010)\citenamefont
  {Urtizberea}, \citenamefont {Natividad}, \citenamefont {Arizaga},
  \citenamefont {Castro},\ and\ \citenamefont
  {Mediano}}]{urtizberea2010specific}%
  \BibitemOpen
  \bibfield  {author} {\bibinfo {author} {\bibfnamefont {A.}~\bibnamefont
  {Urtizberea}}, \bibinfo {author} {\bibfnamefont {E.}~\bibnamefont
  {Natividad}}, \bibinfo {author} {\bibfnamefont {A.}~\bibnamefont {Arizaga}},
  \bibinfo {author} {\bibfnamefont {M.}~\bibnamefont {Castro}}, \ and\ \bibinfo
  {author} {\bibfnamefont {A.}~\bibnamefont {Mediano}},\ }\bibfield  {title}
  {\enquote {\bibinfo {title} {Specific absorption rates and magnetic
  properties of ferrofluids with interaction effects at low concentrations},}\
  }\href {\doibase 10.1021/jp912076f} {\bibfield  {journal} {\bibinfo
  {journal} {J. Phys. Chem. C}\ }\textbf {\bibinfo {volume} {114}},\ \bibinfo
  {pages} {4916--4922} (\bibinfo {year} {2010})}\BibitemShut {NoStop}%
\bibitem [{\citenamefont {Serantes}\ \emph {et~al.}(2010)\citenamefont
  {Serantes}, \citenamefont {Baldomir}, \citenamefont {Martinez-Boubeta},
  \citenamefont {Simeonidis}, \citenamefont {Angelakeris}, \citenamefont
  {Natividad}, \citenamefont {Castro}, \citenamefont {Mediano}, \citenamefont
  {Chen}, \citenamefont {Sanchez}, \citenamefont {Balcells},\ and\
  \citenamefont {Mart\'inez}}]{serantes2010influence}%
  \BibitemOpen
  \bibfield  {author} {\bibinfo {author} {\bibfnamefont {D.}~\bibnamefont
  {Serantes}}, \bibinfo {author} {\bibfnamefont {D.}~\bibnamefont {Baldomir}},
  \bibinfo {author} {\bibfnamefont {C.}~\bibnamefont {Martinez-Boubeta}},
  \bibinfo {author} {\bibfnamefont {K.}~\bibnamefont {Simeonidis}}, \bibinfo
  {author} {\bibfnamefont {M.}~\bibnamefont {Angelakeris}}, \bibinfo {author}
  {\bibfnamefont {E.}~\bibnamefont {Natividad}}, \bibinfo {author}
  {\bibfnamefont {M.}~\bibnamefont {Castro}}, \bibinfo {author} {\bibfnamefont
  {A.}~\bibnamefont {Mediano}}, \bibinfo {author} {\bibfnamefont {D.-X.}\
  \bibnamefont {Chen}}, \bibinfo {author} {\bibfnamefont {A.}~\bibnamefont
  {Sanchez}}, \bibinfo {author} {\bibfnamefont {L.}~\bibnamefont {Balcells}}, \
  and\ \bibinfo {author} {\bibfnamefont {B.}~\bibnamefont {Mart\'inez}},\
  }\bibfield  {title} {\enquote {\bibinfo {title} {Influence of dipolar
  interactions on hyperthermia properties of ferromagnetic particles},}\ }\href
  {\doibase 10.1063/1.3488881} {\bibfield  {journal} {\bibinfo  {journal} {J
  Appl. Phys.}\ }\textbf {\bibinfo {volume} {108}},\ \bibinfo {pages} {073918}
  (\bibinfo {year} {2010})}\BibitemShut {NoStop}%
\end{thebibliography}%

\end{document}